\documentclass[%
 reprint,
nofootinbib,
 amsmath,amssymb,
 aps,
floatfix,
]{revtex4-2}

\usepackage{graphicx}
\usepackage{dcolumn}
\usepackage{bm}
\usepackage[final]{hyperref}
\usepackage{booktabs}
\usepackage{multirow}
\usepackage{placeins}
\usepackage{xcolor}
\usepackage{hyperref}
\usepackage[normalem]{ulem}

\usepackage{lineno}

\begin{document}

\preprint{APS/123-QED}

\title{Feasibility Study to use Neutron Capture for an Ultra-low Energy\\ Nuclear-recoil Calibration in Liquid Xenon}

\author{C.S. Amarasinghe}
\author{R. Coronel}%
\author{D.Q. Huang}
 \email{dqhuang@umich.edu}
\author{Y. Liu}
\author{M. Arthurs}
\author{S. Steinfeld}
\author{W. Lorenzon}
\affiliation{University of Michigan, Randall Laboratory of Physics, Ann Arbor, Michigan 48109, USA}%
\author{R. Gaitskell}
\affiliation{Brown University, Department of Physics, Providence, Rhode Island 02912, USA}%


\date{\today}

\begin{abstract}
The feasibility of  an ultra-low energy nuclear-recoil measurement in liquid xenon using neutron capture is investigated for a small~(sub-kilogram) liquid xenon detector that is optimized for a high scintillation gain, and a pulsed neutron source. The measurement uses the recoil energies imparted to xenon nuclei during the de-excitation process following neutron capture, where promptly emitted $\gamma$ cascades can provide the nuclei with up to $0.3$\,keV$_\text{nr}$ of recoil energy due to conservation of momentum. A successful calibration of scintillation photon and ionization electron yields below this energy will contribute to a greater sensitivity for liquid xenon experiments in searches for light WIMPs.
\end{abstract}

\maketitle

\section{\label{sec:level1}Introduction\protect}

Underground liquid xenon (LXe) time projection chambers~(TPCs) have played an important role in constraining the parameter space available to dark matter in the form of Weakly Interacting Massive Particles (WIMPs) passing through Earth~\cite{aprile2010liquid}.
However, light ($<10$\,GeV) WIMPs are kinematically ill-matched with xenon ($A \approx$ 131) and deposit less energy in the medium than their heavier counterparts.
As a result, dark matter experiments that use LXe suffer a drastic drop in sensitivity for light WIMPs, where the expected signals approach the energy thresholds of the detectors~\cite{lewin1996review}.
Hints of light dark matter in several experiments that use other detector media, like CRESST-II~\cite{angloher2012results}, CDMS-II-Si~\cite{agnese2013silicon}, and CoGeNT~\cite{aalseth2011results}, have therefore stoked interest in characterizing the response of LXe to sub-keV energy depositions.

Matter and radiation deposit energy in LXe by interacting with either atomic electrons, creating electronic recoil (ER) events, or with nuclei, creating nuclear recoil  (NR) events~\cite{davies1994liquid}.
WIMPs are predicted to scatter off nuclei, leaving behind NR signatures~\cite{goodman1985detectability, jungman1996supersymmetric}. 
Both ER and NR events create detectable scintillation photons (S1) and ionized electrons, with some energy being lost as heat~\cite{aprile2010liquid}.
In a dual-phase LXe TPC\footnote{The basic operating principle of a typical dual-phase LXe TPC is described in Section III. E of Ref.~\cite{aprile2010liquid}.}, the ionized electrons are drifted towards and extracted into a gaseous xenon space by an electric field, where a secondary larger flash of light (S2) is produced by electroluminescence. 
The ratio S2/S1 is smaller for an NR than for an ER, a feature of LXe that allows ER events to be rejected with high efficiency ($>$ 99\% at 50\% NR acceptance) from potential WIMP-induced NR events~\cite{akimov2021two, collaboration2018dark, akerib2020discrimination}.

For a particular experiment to infer the WIMP mass and interaction cross section in case of a discovery, a map from the space of observed $\{ \text{S1}, \text{S2} \}$ signals to NR energy is required.  
The production of S1 and S2 signals in LXe due to NR events of known energies has been characterized in a series of measurements over the past two decades~\cite{aprile2006simultaneous, sorensen2009scintillation, manzur2010scintillation, horn2011nuclear, aprile2013response}.
As a result, a detector-independent picture of how LXe produces photons and electrons in response to NR events has emerged.
In recent years there has been a concerted effort to determine these quanta yields at lower energies, allowing experiments to be sensitive to lighter WIMPs~\cite{lin2015scintillation, akerib2016low, aprile2018simultaneous, lenardo2019measurement, lux2022improved}.
The current lowest energy measurements have found 1.1-1.3 ionized electrons per 0.3\,keV$_\text{nr}$ recoil~\cite{lux2022improved, lenardo2019low}, and 1.3 scintillation photons per 0.45\,keV$_\text{nr}$ recoil~\cite{lux2022improved}.
This work presents an experimental concept to measure these yields below 0.3 keV$_\text{nr}$.

Previous measurements of the photon and electron yields in LXe have used the elastic scattering of neutrons as a source of nuclear recoils. 
We propose to use xenon nuclei that have captured neutrons.
The nuclear recoils of interest are generated by the asymmetric emission of de-excitation $\gamma$ cascades that leave the TPC undetected, as suggested in Ref.~\cite{dqthesis}.
The idea of using neutron capture to access low recoil energies was implemented for germanium in Ref.~\cite{jones1975energy}, and has been repeatedly studied in that material~\cite{collar2021germanium, agnese2016new, thulliez2021calibration}.
Here we introduce a technique to implement this idea in LXe. 

The details of this work correspond to simulations carried out for the Michigan Xenon (MiX) detector, a small (400g active volume) dual-phase LXe TPC with an excellent light collection efficiency and energy resolution~\cite{stephenson2015mix}, although the principles apply to any small TPC.
A pulsed neutron source and a neutron moderator surrounding the detector are assumed for the simulation.
These components are found to be crucial in creating a collection of neutron captures in each pulse that are unaccompanied in time by other sources of NR, in addition to reducing backgrounds from spurious electron emissions.

The paper is organized as follows. 
Section~\ref{sec: strategy} describes how the neutron capture induced nuclear recoils are selected, while in section~\ref{simsSection} we present details about the Monte Carlo simulation. 
In section~\ref{SIGNAL}, we report how to optimize the neutron capture signal by varying aspects of the experimental setup.
Background and pile-up events are discussed in section~\ref{Background}, along with changes to the setup required to minimize them.
Section~\ref{discussionSection} describes the implications this measurement could have on the sensitivities for light WIMP searches.
We conclude in section~\ref{conclusion}.

\section{General Approach} \label{sec: strategy}

After a xenon nucleus captures a neutron, the $\gamma$ cascade leaves it with up to 0.3\,keV$_\text{nr}$ of kinetic energy that it dissipates among neighboring atoms, producing photons and electrons.
In order for the associated S1 and S2 signals to be cleanly recorded by a data acquisition system, the acquisition window cannot be contaminated by other ER or NR events. 
Only acquisition windows free of ER events are chosen for the measurement, by selecting captures in which the de-excitation $\gamma$ cascade escapes the active volume, and also by rejecting events with ER events that originate externally.
A neutron capture event can be positively identified if a separate detector outside the TPC detects the $\gamma$ cascade, providing a timestamp to tag the capture NR. 

Using a pulsed neutron source and a thin moderator between the source and detector, a set of neutron capture events suitable for the measurement can be produced in each pulse.
Since the neutron capture cross section is roughly proportional to the inverse speed of the incident neutron (except at resonances), capture events are mostly caused by slow neutrons in the TPC.
The role of the moderator is to slow down monoenergetic neutrons from the source, while discouraging neutron capture in the moderator itself, as the resulting $\gamma$ rays are a source of pile-up.
Accordingly, the simulation shows that partial neutron moderation is ideal.
The thin moderator allows fast neutrons into the TPC first, which are likely to scatter, followed by slower neutrons that are captured.
In this arrangement NR events due to neutron capture can be isolated from scattering events with an appropriate time cut.

The observed S1 and S2 pulses have to be associated with the energy of the nuclear recoil that produced them.
While many previous measurements have had precise knowledge of the recoil energies, for example by using the angle of the scattered neutrons~\cite{akerib2016low, lenardo2019measurement}, this measurement relies on a model of energy deposition in LXe due to the neutron capture process.
The distribution of NR energies simulated by this model will be used to calculate the sizes of the S1 and S2 signals according to parameterized estimates of the yields below $0.3$ keV$_\text{nr}$.
These parameters can be adjusted to fit the calculated S1 and S2 sizes to the observed data, as performed in Ref.~\cite{dqthesis}.
The energy deposition model and its uncertainty are presented in Section~\ref{simsSection} and discussed in detail in Appendix~\ref{app:A}.

\subsection{Neutron Interactions in Liquid Xenon} \label{subsec nCap}

Upon capturing a neutron, most xenon isotopes promptly de-excite (within 1 ns) to their ground state by releasing a cascade of $\gamma$ rays: $^A$Xe + n $\rightarrow ^{A+1}$Xe + $\sum \gamma$~\cite{molnar2004handbook}. In some cases, this process also releases internal conversion electrons. If the de-excitation transition is direct, that is if a single $\gamma$ ray carries away all the excitation energy (or equivalently, if several $\gamma$ rays are emitted in the same direction), a nucleus initially at rest is given the maximum recoil energy
\begin{equation}
E_{R, \text{max}} = \frac{S^2_n}{2 M_\text{Xe}} \approx S_n^2 \left( \frac{4 \times 10^{-6}}{\text{MeV}} \right),
\end{equation}
where $S_n$ is the neutron separation energy of the newly created xenon isotope $^{A+1}$Xe, and $M_\text{Xe}$ is its mass.
Table~\ref{tab:nGamma} shows neutron separation energies and the corresponding maximum recoil energies for each of the naturally occurring xenon isotopes.

\begin{table*}
    \centering
    \begin{ruledtabular}
    \begin{tabular}{c c c c | c c c}
        Target Isotope & Abundance (\%) & $E^*_1$ (keV) & Capture Cross Section (b) & Product Isotope & $S_n$ (keV) & $E_\text{R, \text{max}}$ (keV$_\text{nr}$) \\
        \midrule
        $^{124}$Xe & 0.1  & 354.0   & 165 $\pm$ 20    & $^{125}$Xe & 7603 & 0.230 \\
        $^{126}$Xe & 0.1  & 388.6  & 3.8 $\pm$ 0.5   & $^{127}$Xe & 7223 & - \\
        $^{128}$Xe & 1.9  & 422.9  & 5.2 $\pm$ 1.3   & $^{129}$Xe & 6908 & 0.187 \\
        $^{129}$Xe & 26.4 & 39.6  & 21 $\pm$ 5      & $^{130}$Xe & 9256 & 0.332 \\
        $^{130}$Xe & 4.1  & 536.1  & 4.8 $\pm$ 1.2   & $^{131}$Xe & 6605 & 0.168 \\
        $^{131}$Xe & 21.2 & 80.2   & 85 $\pm$ 10     & $^{132}$Xe & 8937 & 0.305 \\
        $^{132}$Xe & 26.9 & 667.7  & 0.42 $\pm$ 0.05 & $^{133}$Xe & 6440 & - \\
        $^{134}$Xe & 10.4 & 847.0  & 0.27 $\pm$ 0.02 & $^{135}$Xe & 8548 & - \\
        $^{136}$Xe & 8.9  & 1313.0 & 0.26 $\pm$ 0.02 & $^{137}$Xe & 4025 & 0.060 \\
    \end{tabular}
    \end{ruledtabular}
    \caption{Properties of xenon nuclei that are relevant to interactions with slow neutrons: natural abundances~\cite{Gammabook}, energies of the first excited nuclear state $E^*_1$~\cite{sukhoruchkin2012excited}, thermal neutron capture cross sections, neutron separation energies $S_n$ of the product nuclei, and the maximum recoil energy $E_\text{R, \text{max}}$ imparted to the product nuclei by the $\gamma$ cascades following capture~\cite{Gammabook}. 
    Of primary interest to the proposed measurement are $^{129}$Xe and $^{131}$Xe due to their large natural abundances, large thermal neutron capture cross sections, and the prompt $\gamma$ cascades of their capture products.
    The isotopes with missing data in the last column produce activated products upon neutron capture that do not decay promptly.}
    \label{tab:nGamma}
\end{table*} 

{Most de-excitations occur with the emission of several $\gamma$ rays that exit the nucleus in different directions, leaving the nucleus with recoil energy $E_R \leq E_{R, \text{max}}$. 
As a result, the recoil spectra of each isotope will be a distribution bounded from above by $E_{R, \text{max}}$, assuming the momentum transferred to the nuclei from the collision with the neutrons is negligible.
Otherwise one has to add to this bound the energy transferred to the nucleus from the collision of approximately $E_\text{n}/131$, where $E_\text{n}$ is the kinetic energy of the neutron when it was captured.
The NR events selected for this measurement are produced from the capture of neutrons with an average energy of 20\,eV, which results in a negligible $0.15$\,eV$_\text{nr}$ contribution to the recoil energy.}
In contrast to studies performed with germanium detectors, where monoenergetic recoils of 0.245 keV$_\text{nr}$ were tagged using a $\gamma$ ray from a low energy excited state of $^{73}$Ge~\cite{collar2021germanium}, the entire distribution of capture-induced recoils in xenon will be used.

{Metastable states with lifetimes many orders of magnitude greater than the capture states can be populated by neutron capture or by the inelastic scattering of neutrons by xenon.
The most abundantly created metastable states are $^{129\text{m}}$Xe and $^{131\text{m}}$Xe, which produce prominent 236 keV and 164 keV $\gamma$ lines, respectively~\cite{ni2007preparation}. 
These $\gamma$ rays also recoil xenon nuclei, but the resulting events do not contribute to the NR calibration for two reasons.
Most importantly, the magnitude of nuclear recoils caused by the emission of these $\gamma$ rays is $\mathcal{O}$(0.1\,eV$_\text{nr}$), and will not be sufficient to produce quanta.
Second, the half lives of the metastable states are too long (seconds to days) for them to be selected in time along with the neutron captures in each pulse.}

Elastic scattering is an inefficient process to transfer energy from slow neutrons to xenon nuclei due to the large difference in their masses and because no energy goes into altering nuclear states~\cite{knoll2010radiation}.
Due to this inefficiency, a neutron has to scatter numerous times before it is captured, resulting in a high rate of elastic scatters immediately after the neutron pulse.
The simulations show that the thermalization time of neutrons in LXe is $\mathcal{O}$($10\,\mu$s), after which they are readily captured.
The average time it takes for a neutron to be captured in the TPC after being emitted is $\mathcal{O}$($100\,\mu$s).
This is long enough for the time cut to be effective in isolating a collection of NR events produced only by captures, with an acceptance of around 80\% for the experimental configuration discussed in Section~\ref{sec:apparatus}.

\subsection{Signal Selection} \label{subsec: signal selection}

\begin{figure}
    \includegraphics[width=0.45\textwidth]{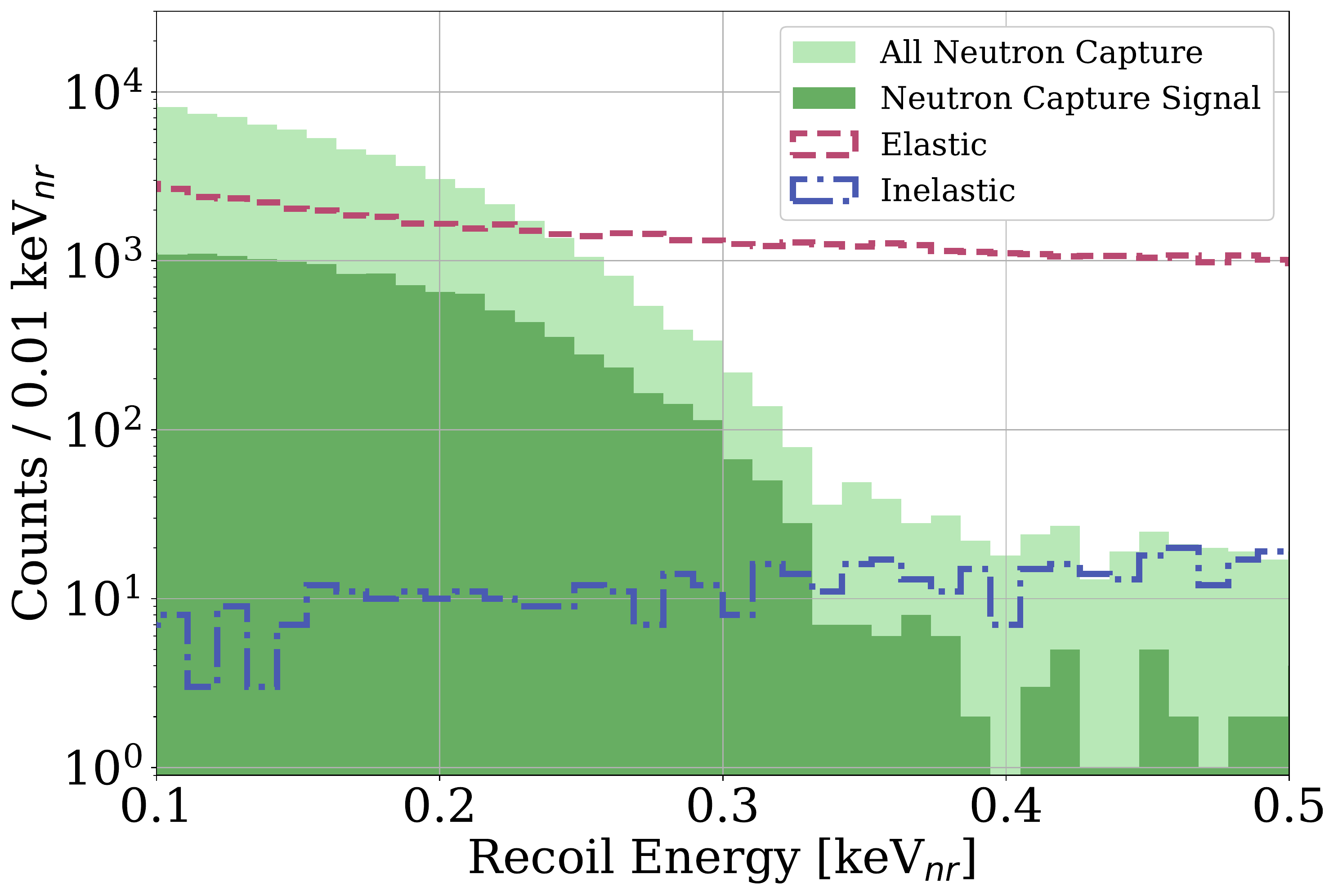}
    \caption{Nuclear recoil spectrum due to neutron interactions simulated in the MiX detector. 
    The shaded light green histogram (140,000 counts) shows all recoil events due to neutron captures, while the shaded dark green portion (20,000 counts) only retains those where all of the $\gamma$-rays from the nuclear de-excitation process escape the active volume.
    The concentration of signals below about 0.3\,keV$_\text{nr}$ provides an opportunity for a measurement of quanta in this energy region.
    Also shown are the recoil events due to elastic (dashed magenta) and inelastic (dashed-dot blue) neutron scatters. 
    All inelastic scatters are shown, regardless of whether their de-excitation $\gamma$-products escape the TPC.}
    \label{fig:n capture}
\end{figure}

Although all capture events result in a recoiling nucleus, the signal events are defined to be neutron captures that did not deposit more than 10 eV$_\text{nr}$ before capture in the TPC, and where the entire $\gamma$ cascade escapes the TPC without depositing energy in it. 
If internal conversion electrons and subsequent atomic emissions (X-rays and Auger electrons) are produced in an event, it is discarded. 
This ensures that signal events have a pure NR signature.
An example of an NR spectrum due to the neutron capture de-excitation process, and the subset of signal events, is shown in Fig.~\ref{fig:n capture} for a detector geometry that is discussed in Section~\ref{simsSection}.
Also shown are the low energy nuclear recoil events due to elastic and inelastic neutron scattering.
At energies below $0.3$\,keV$_\text{nr}$, neutron capture events contribute to the majority of the NR spectrum, and signal events make up around 15\% of captures in the TPC. 
The capture-induced NR events above around 0.3\,keV$_\text{nr}$ are due to collisions with faster moving neutrons, and can be removed with a time cut.

\subsection{Tagging signal events using the LXe skin}

Signal events can be positively identified if their $\gamma$ cascades are detected outside the TPC.
A natural location to detect the interactions of these $\gamma$ rays is the detector skin, the volume of LXe immediately outside the TPC.
In the MiX detector, this volume is ideal for tagging signal events due to its large size.
Simulations show that 70\% of signal events can be tagged using an instrumented skin with a 100\,keV$_\text{ee}$ energy threshold. 
In other words, less than 30\% of signal events emit $\gamma$ cascades that escape not only the TPC, but the surrounding skin region as well.
No significant bias on the NR energy spectrum is observed when taggable signals are selected. 
Tagging offers a major reduction in the single-electron background commonly observed in LXe TPCs that may otherwise dominate the number of events from neutron capture that also produce single electrons~\cite{akerib2020investigation}.

\section{Simulation} \label{simsSection}

The NR energy distribution of neutron capture events depends on the detector's neutron environment and the nuclear properties of xenon.
The passage of neutrons emitted from an external source through the MiX detector, and the energy deposits of neutron capture events were studied using a Monte Carlo simulation built with the GEANT4-based application BACCARAT, a detector independent framework developed by the LUX and LZ collaborations~\cite{akerib2012luxsim, akerib2020simulations}. 
The MiX detector geometry was tessellated from existing CAD drawings and imported into this framework using the CADMesh package~\cite{poole2012acad}.

\subsection{Neutron model}

\begin{figure}
    \includegraphics[width=0.48\textwidth]{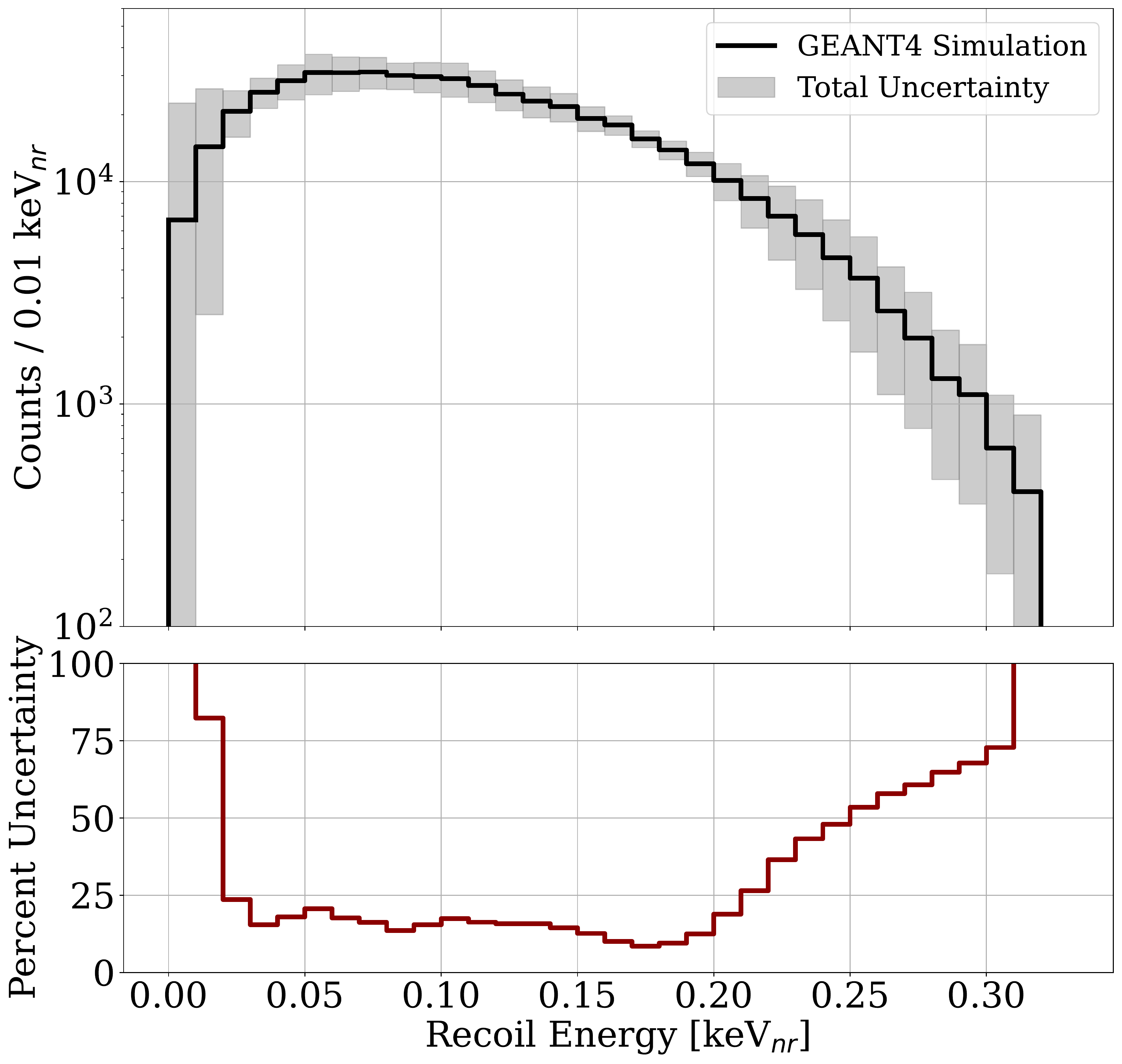}
    \caption{\emph{Top:} NR spectrum due to thermal neutron capture in LXe simulated using GEANT4. The gray uncertainty band represents the total uncertainty, which incorporates discrepancies in the $\gamma$ spectra between the ENSDF and the EGAF files.
    \emph{Bottom:} The error band in the top panel is presented as percent uncertainty for clarity.
    }
    \label{fig:nCapError}
\end{figure}

The low energy neutron transport processes are modeled using the QGSP\_BIC\_HP physics list in GEANT4, and the de-excitation process following neutron capture is simulated with the GEANT4 photon evaporation model.
The photon evaporation model simulates discrete and continuous $\gamma$ cascades using the Evaluated Nuclear Structure Data File (ENSDF), and also simulates internal conversion electrons~\cite{allison2016recent}.
The photon evaporation algorithm conserves energy and momentum, and appears to handle the dynamics of cascade production sufficiently well, although it has not been experimentally validated.
Validation requires measurements of the $\gamma$ spectra for each multiplicity\footnote{Multiplicity refers to the number of $\gamma$ rays emitted in a de-excitation.} that have so far only been made for the target isotope $^{136}$Xe~\cite{albert2016measurement}.
Since the experimental concept relies on a comparison with simulations, a custom algorithm was implemented to generate nuclear recoils from neutron capture and used to calculate the uncertainty of the NR energy spectrum.
This uncertainty, {shown in Fig.~\ref{fig:nCapError}, incorporates discrepancies in the $\gamma$ spectra between ENSDF and the evaluated gamma ray activation file (EGAF), which is an experimental database of multiplicity-independent neutron capture $\gamma$ energies~\cite{EGAFdata}}.
The uncertainty calculation is discussed in Appendix~\ref{app:A}.

$^{136}$Xe is not important to the proposed measurement due its low neutron capture cross section, comprising only 0.1\% of neutron capture events in natural xenon, and because its largest recoil energy is 60\,eV$_\text{nr}$, which is too small to produce a signal.
However it is the only isotope for which data exists to make a comparison with GEANT4 that properly takes into account $\gamma$ spectra at each multiplicity. 
Using the custom algorithm that generates nuclear recoil events from neutron capture, the prediction of the NR spectrum from those data was compared with GEANT4 and a weighted average difference of 39.7\% was found in the $0-60\,$ eV$_\text{nr}$ range.
In Appendix~\ref{app:A} this calculation is presented and it is argued why such large discrepancies are not expected for the other isotopes of xenon if measurements of their $\gamma$ spectra are eventually made. 

\subsection{Description of the setup}\label{sec:apparatus}

\begin{figure}
    \includegraphics[width=0.47\textwidth]{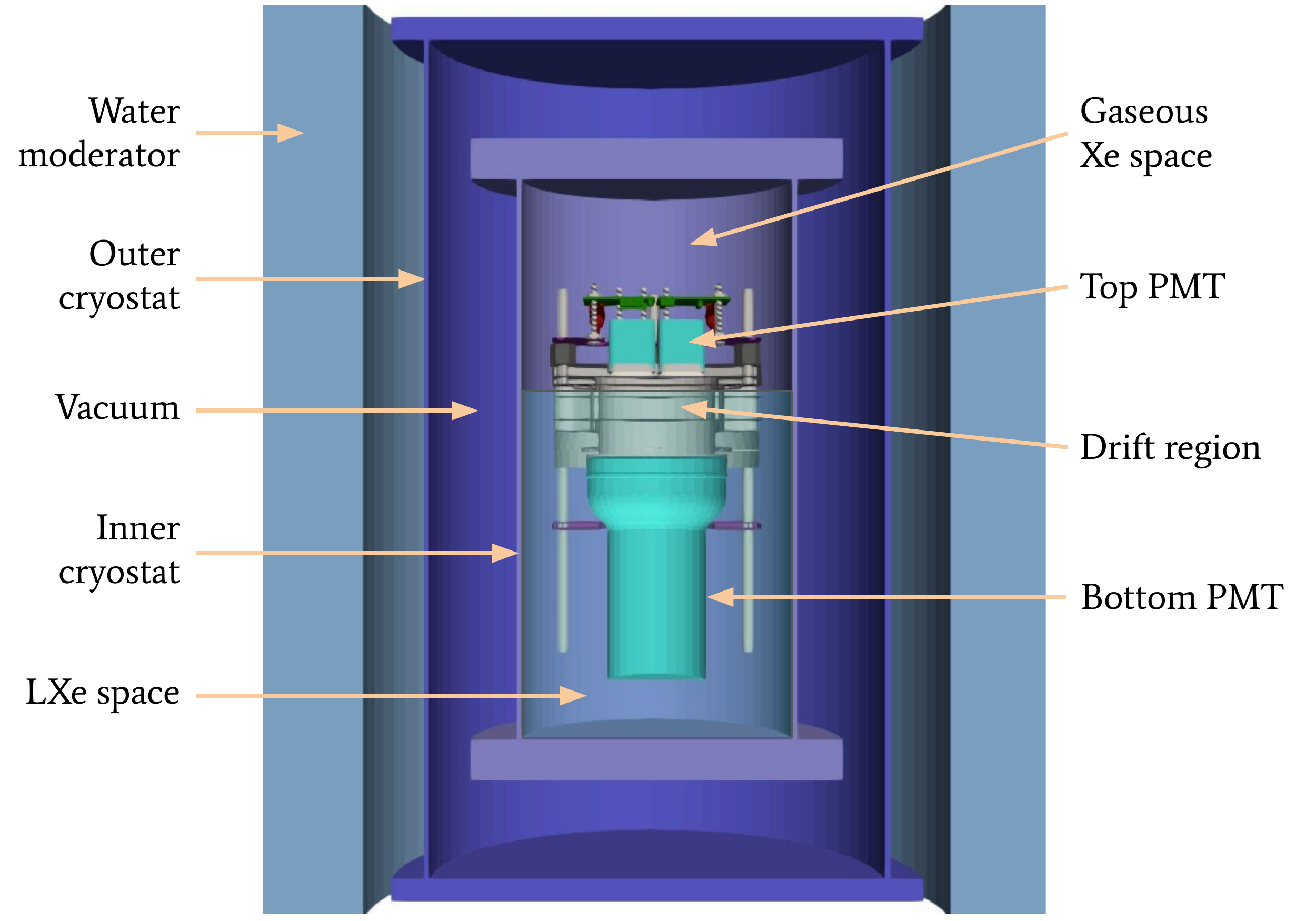}
    \caption{3D model of the MiX detector. 
    The inner cryostat encloses the LXe space that partially submerges the TPC assembly, and thus the TPC contains only a small fraction of the LXe in the system. The thickness of the water tank shown here is 5\,cm.
    }
    \label{fig:MiX}
\end{figure}

The MiX detector is a small dual-phase TPC at the University of Michigan that is ideally suited to study properties of LXe.
A cross section of the detector is shown in Fig.~\ref{fig:MiX}.
The MiX detector has a drift chamber with a diameter of 62.5\,mm and a height of 12\,mm.
It was designed and built to have good signal gains, with scintillation and ionization gains of (0.239 $\pm$ 0.012) photoelectrons/photon and (16.1 $\pm$ 0.6) photoelectrons/electron, respectively~\cite{stephenson2015mix}.
The high scintillation gain, which is crucial to measure the LXe response to low energy interactions~\cite{aprile2010liquid}, is more than a factor of 2 larger than that of typical $\mathcal{O}$(100\,kg) scale detectors. 
This makes the MiX detector a suitable candidate to perform an ultra-low energy NR calibration in LXe.

The feasibility study assumes a $2.45$\,MeV monoenergetic neutron source, modeled after an upgraded Adelphi Technologies DD109 Deuterium-Deuterium (D-D) neutron generator.
The source has the ability to create pulses as short as 20\,$\mu$s at an instantaneous rate of $10^9$\,n/s~\cite{dqthesis}.
A point source of neutrons that originates one meter\footnote{The conclusions of this study do not strongly depend on this distance.} away from the center of the TPC is simulated.
The solid angle of the neutrons that intercept the setup ranges from 0.1 to 0.35 steradians, depending on the size of the water tank. 
The ability to produce short pulses of neutrons is essential to isolate neutron capture events and mitigate single-electron background events, as discussed in Sections~\ref{SIGNAL} and~\ref{Background}, respectively.
It is required that neutron interactions following a pulse completely die off before the next pulse starts. 
This ensures that the timing effects in each cycle can be treated independently. 

\begin{figure}
    \centering
    \includegraphics[width=0.45\textwidth]{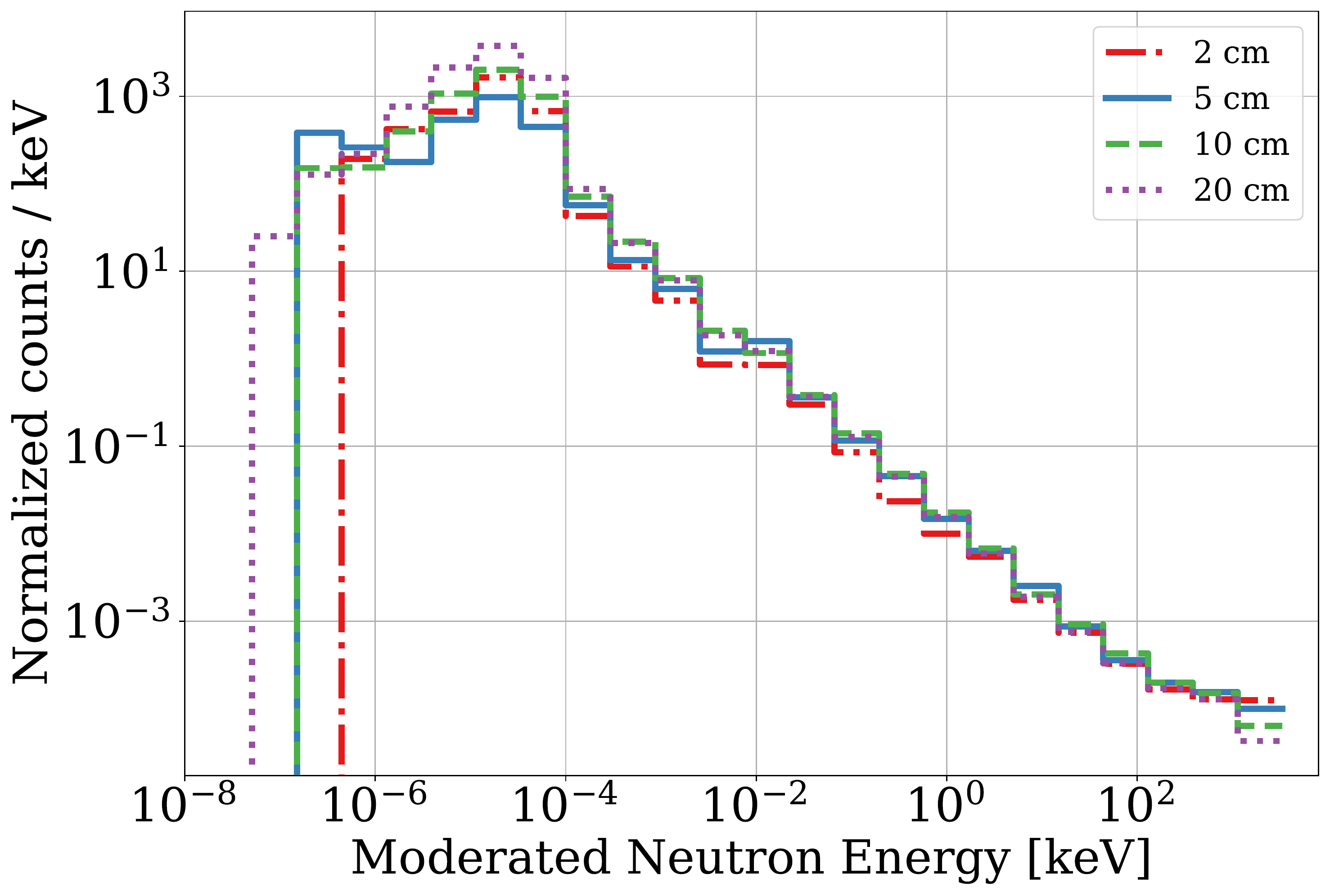}
    \caption{Kinetic energy distributions of neutrons as they enter the TPC after being moderated by the water tank, shown for various thicknesses of the tank.
    }
    \label{fig:internalNeutronEnergy}
\end{figure}

\begin{figure}
    \centering
    \includegraphics[width=0.45\textwidth]{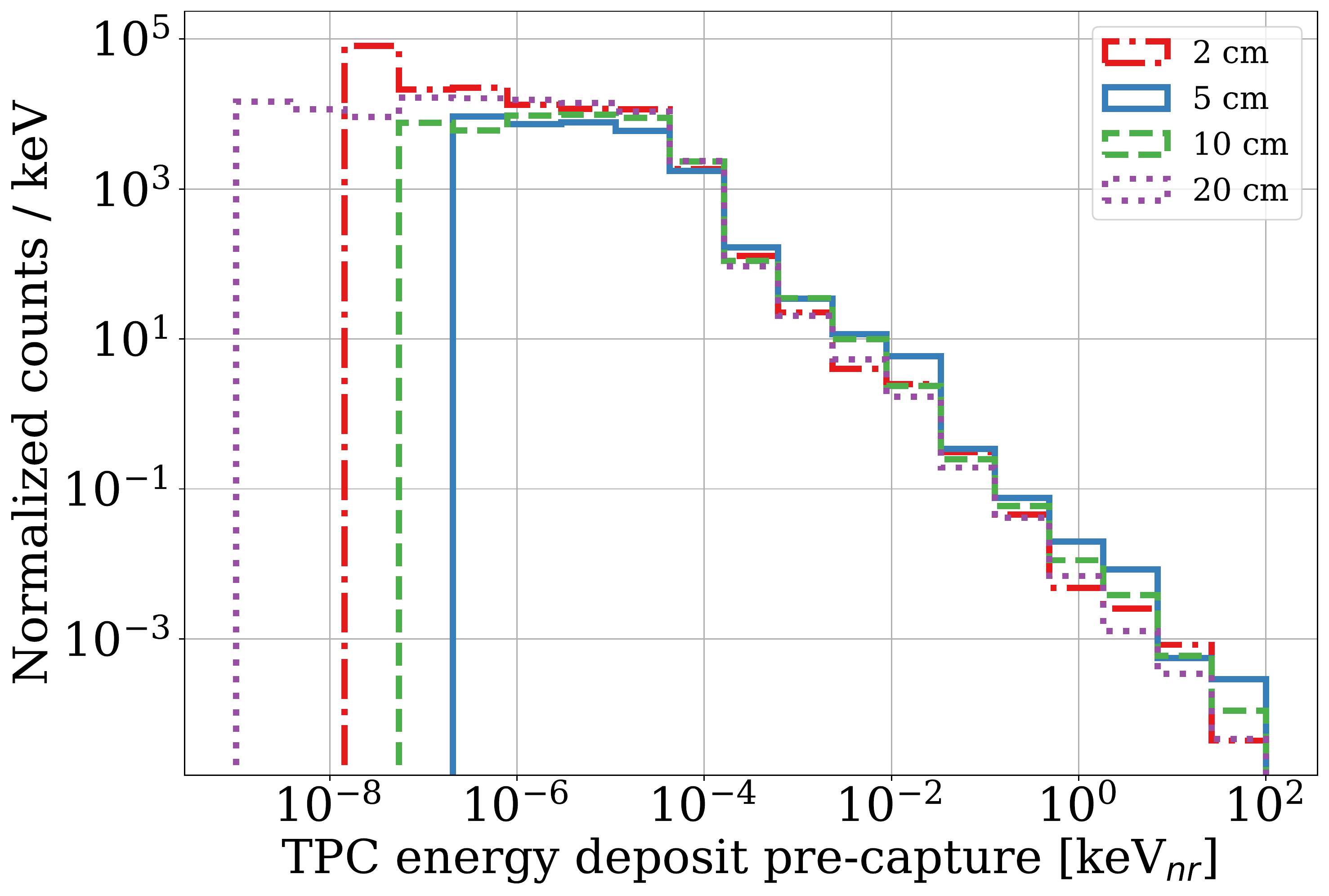}
    \caption{Distributions of the NR energy transferred to xenon in the TPC by neutron scattering before capture, shown for various thicknesses of the water tank.}
    \label{fig:precaptureDep}
\end{figure}

{
A cylindrical water tank surrounds the detector to moderate the D-D neutrons for capture. 
Neutron kinetic energies are shown in Fig.~\ref{fig:internalNeutronEnergy} as they enter the TPC for various tank thicknesses.
The neutrons are further moderated by xenon in the TPC. 
The NR energy distributions for neutrons prior to capture are shown in Fig.~\ref{fig:precaptureDep} for various tank thicknesses.
For signal events surviving the time cut, discussed in Section~\ref{sec:timing}, the energy dissipated in the TPC before the neutrons are captured is insufficient to produce quanta. 
The simulation shows that 90\% of those signals events are due to neutrons that deposit less than $6 \times 10^{-5}$\,keV$_\text{nr}$ in the TPC by scattering.
}

\begin{figure}
    \centering
    \includegraphics[width=0.49\textwidth]{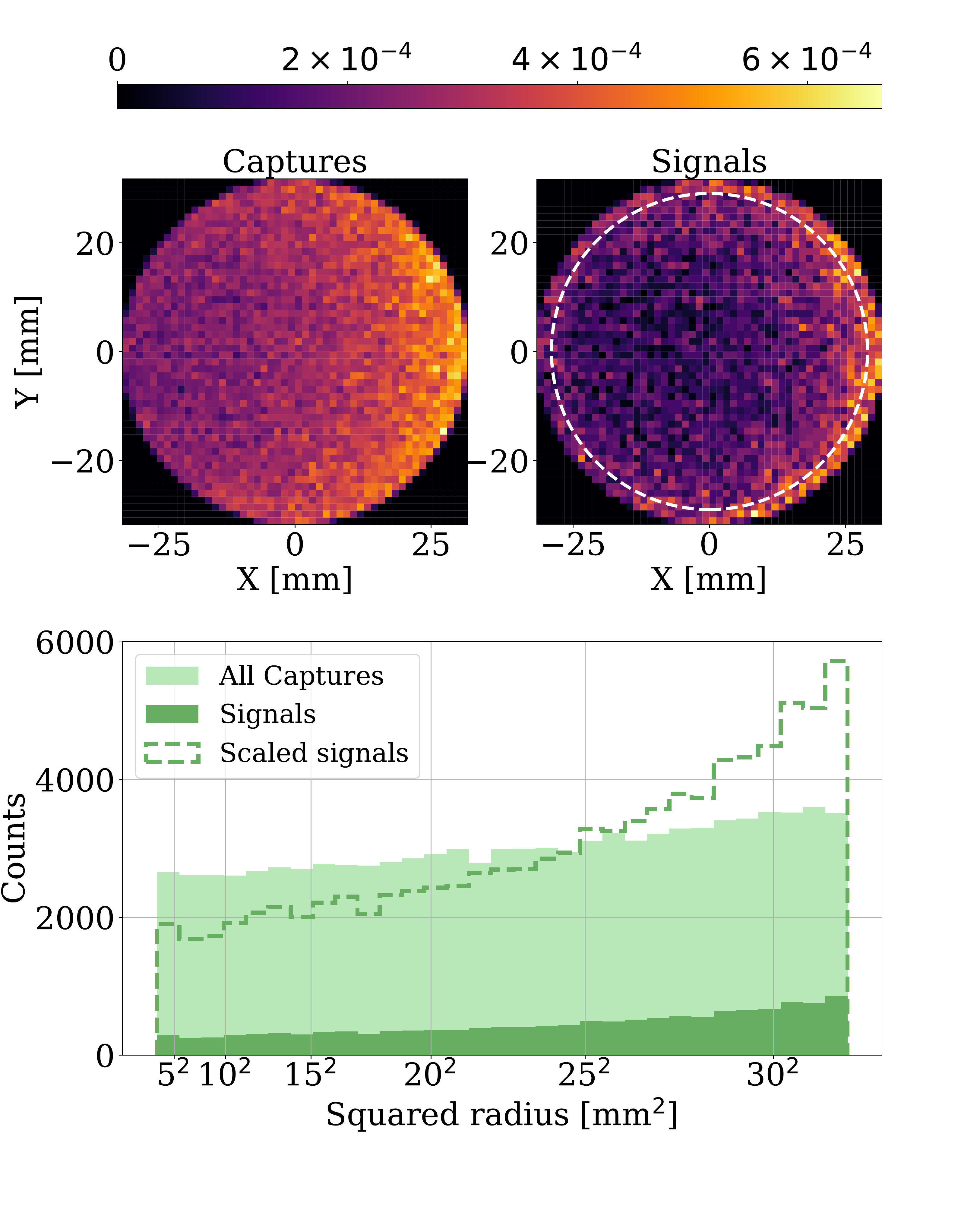}
    \caption{\emph{Top:} Neutron capture locations in the TPC (left), and signal event locations (right), each normalized to unity.
    The neutrons enter the water tank from the right, which causes the higher concentration of captures on the right edge.
    The white circle on the right plot indicates the fiducial radius defined in the MiX detector~\cite{stephenson2015mix}.
    Only 20\% of signal events fall outside its radius.
    \emph{Bottom:} Radial positions of capture (light green) and signal (dark green) events.
    The signal population scaled to the total counts of captures is also shown (dashed) to demonstrate the higher concentration of signal events near the walls of the TPC.
    }
    \label{fig:locs}
\end{figure}

The radial profiles of neutron capture interactions in the active volume of the MiX detector are shown in Fig.~\ref{fig:locs}.
Neutron capture events are concentrated on the edge of the TPC closest to the neutron source.
The signal population is also largely near that edge, because there is a geometric advantage for $\gamma$ cascades escaping the TPC near a wall.
Although the fiducial volume in the MiX detector is only well defined within a radius of 29\,mm, 80\% of the signal events are retained~\cite{stephenson2015mix}. 

\section{Signal Optimization} \label{SIGNAL}

{
There are two factors that contribute to an optimal signal: a high yield of signal events, and the separability of these events from neutron scattering events. 
The simulation shows that the presence of a water tank to moderate the D-D neutrons boosts the fraction of signal events, and that its thickness can be tuned to gain a favorable separation of neutron capture events.
The analysis of the time distributions of TPC neutron capture events provides suitable values for the neutron generator pulse width $w_n$ and pulsing frequency $f$.
However, $w_n$ and $f$ are more strongly constrained by the rates of background and pile-up events.
As discussed in Sections~\ref{sec:timing} and~\ref{subsec: HE-ER}, the optimal parameters for this experiment are a water tank of thickness 5\,cm, neutron pulse width $w_n = 30\,\mu$s, and pulsing frequency $f = 60$\,Hz.
}

\subsection{Signal and Target Energy Estimates}
Following the optimal configuration presented in Section~\ref{SIGNAL}, estimates for the signal event rate and target energy are discussed.
With an instantaneous rate of $10^9$ n/s emitted isotropically, 330 neutrons enter the water tank in each pulse. 
Of these, roughly 0.1 neutrons (0.03\%) are captured in the TPC, but only 0.015 events (15\% of neutron capture events) end up as signal events.
After applying position cuts that only keep events within the MiX fiducial volume, i.e. within a 29\,mm radius of the active region, 0.004 signal events per pulse survive.
This results in a signal event every 250 pulses, or roughly 1 signal every 4 seconds at a pulsing frequency of 60\,Hz.
Using an instrumented LXe skin with an estimated capture tagging efficiency of 70\%, a final rate of 0.2 usable signal events per second is expected.

The lowest NR energy for this experimental configuration depends on the detector's intrinsic and neutron-induced backgrounds, the exposure, and the scintillation and ionization yields.
Even a basic estimate of this target NR energy requires an assumption of the yields below 0.3 keV$_\text{nr}$ where there is currently no data, in addition to assumptions about the yet unmeasured background levels in the MiX detector.
The Noble Element Simulation Technique (NEST) v2.0.1 NR yield model (which was modified to remove the sharp cutoff in the yields at 0.2\,keV$_\text{nr}$) and the photon evaporation model were used to simulate quanta produced by the neutron capture events. 
The simulation predicts a drop in quanta production at 0.13\,keV$_\text{nr}$, where an average of 0.2 ionized electrons are expected. 
The drop in quanta production was confirmed by weighting the yields directly from NEST with the NR spectrum.
The quanta simulated for two months of runtime ($10^6$ usable signal events) were compared with the weighted NEST yields using a $\chi^2$ test, scanned over various energy thresholds. 
The threshold energy at which the goodness of the fit stopped improving is consistent with the 0.13\,keV$_\text{nr}$ target energy.
The NEST extrapolation predicts an average of 0.2 ionized electrons at 0.13\,keV$_\text{nr}$.
Thus this energy threshold is within reach of a two-month run. 

\subsection{Timing of Neutron Interactions} \label{sec:timing}

\begin{figure}
    \centering
    \includegraphics[width=0.45\textwidth]{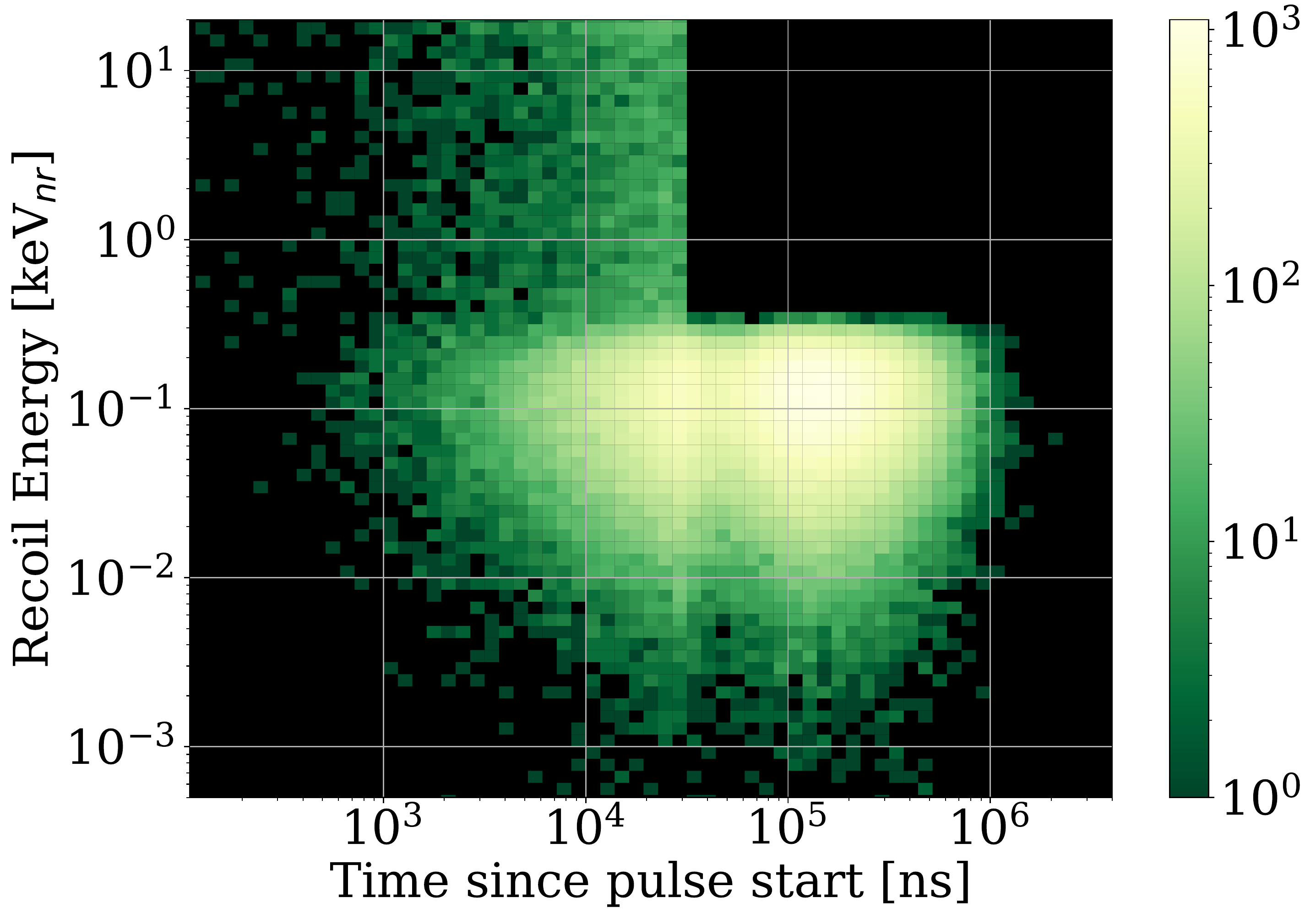}
    \caption{Recoil energy of the xenon atoms at the time the neutrons were captured for $10^5$ neutron captures. 
    The simulation corresponds to a 5\,cm water tank and a 30\,$\mu$s pulse width.
    A small fraction of neutrons, shown in the top left quadrant, reaches the TPC early with enough energy to cause collisional recoil energies noticeably greater than the 0.3\,keV$_\text{nr}$ possible by the $\gamma$ cascades alone.
    The time of flight of these events is $\mathcal{O}$(100\,ns), so they abruptly cease shortly after the pulse ends at 30\,$\mu$s.
    }
    \label{fig:recoilvFlight}
\end{figure}

Most neutrons are captured between 10\,$\mu$s and 1\,ms after they are emitted by the source as shown in Fig.~\ref{fig:recoilvFlight} for a 5\,cm water tank and a 30\,$\mu$s pulse width.
This is due to the joint effect of the neutrons spending most of this time losing energy in the moderator tank and the fact that neutron capture cross sections scale inversely to the incident neutron speed~\cite{molnar2004handbook}.
The \emph{signal window} corresponding to a pulse is defined as the period of time that starts when all neutron scattering has died off, and ends when 99\% of neutron capture signal events have been produced.
The typical size of a signal window for a 5\,cm thick water tank is 0.55\,ms.

Less than 2\% of the captured neutrons are not slowed down significantly and are captured early in the TPC, creating extra recoil energy due to the collision.
This population can be seen in the top left quadrant of Fig.~\ref{fig:recoilvFlight} with recoil energies greater than $0.3$ keV$_\text{nr}$, which is the maximum recoil energy expected due to $\gamma$ emissions from stationary xenon nuclei.
These events extend up to the width of the pulse, and can be easily removed with a time cut.

\begin{figure}
    \centering
    \includegraphics[width=0.48\textwidth]{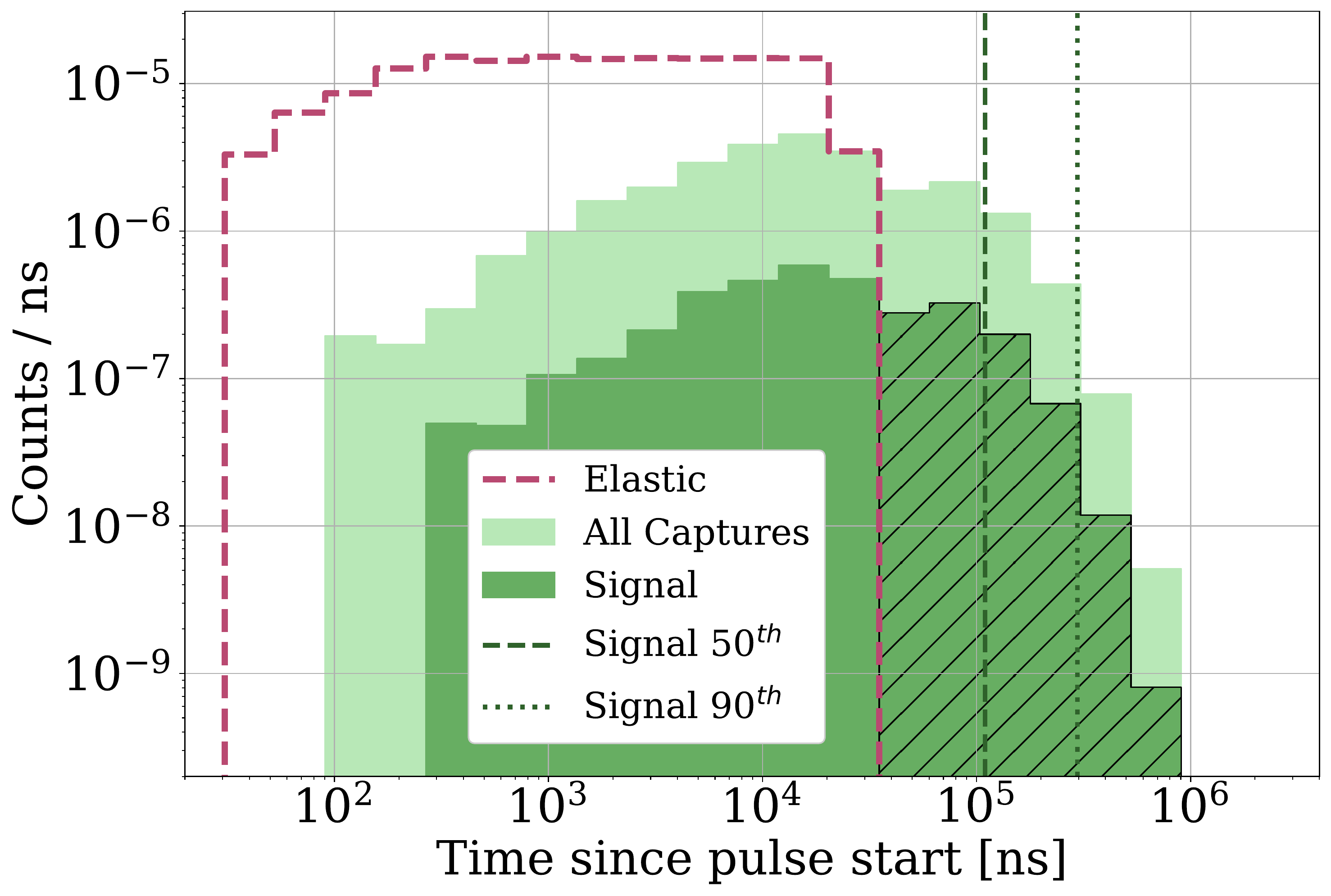}
    \caption{Time distribution of the neutrons that interact with the active LXe volume in the TPC, from a simulation done for a 5\,cm water tank and $w_n = 30\,\mu$s.
    The total counts due to neutron capture (light green) and elastic scattering (dashed magenta) are normalized to unity.
    Inelastic scattering events are omitted from this plot for clarity as their rate is a hundred-fold less than the elastic rate.
    All events shown here deposit less than $1$ keV$_\text{nr}$.
    The dark green histogram shows all signal events, and the hatched portion shows the signal events that occur after the last scattering time.
    Visual checkpoints for when $50\%$ and $90\%$ of all signal events occur are shown with the vertical dashed and dotted lines, respectively.
    }
    \label{fig:timeStructure}
\end{figure}

An advantage of using a water tank moderator together with a pulsed neutron source is that the former's thermalization effect separates the scattering from capture events in time, creating a pure collection of neutron capture events over several pulses.
The signal separability $\text{T}_\text{NR}$ is quantified (averaged over numerous cycles) as
\begin{align}
\label{eq:SR_NR}
    &\text{T}_\text{NR} (E_{\downarrow}, w_n) = \nonumber \\  
    &\frac{\text{Number of capture signals after the last scatter}}{\text{Number of scattering events}},
\end{align}
where only events with deposited energy below $E_\downarrow$ are kept, and where the neutron pulse width is $w_n$. 
The time at which all the scattering interactions have died off is defined as the last scatter time.
Therefore, the numerator represents the signal events in a cycle that are desirable for the measurement since they will not be accompanied by recoil events due to scattering.
Figure~\ref{fig:timeStructure} shows these populations for $E_\downarrow = 1$\,keV, a 5\,cm water tank, and a pulse width $w_n = 30\,\mu$s.
The area of the hatched portion represents the numerator of Eq.~\ref{eq:SR_NR} while the area of the elastic scatter portion (under the dashed magenta line) represents the denominator.
Although a neutron pulse width of $w_n = 30\,\mu$s is used for Fig.~\ref{fig:timeStructure}, the timing of neutron capture events is not very sensitive to $w_n$.
Rather, it is the timing of the neutron scattering processes, which take place in less than 50\,$\mu$s, that more keenly depend on $w_n$.
This fact can be used to maximize the number of signal events that occur after the last scatter time. 
The effects of varying the pulse width are discussed in Section~\ref{ss: pulseWidth}.

\begin{figure}
    \centering
    \includegraphics[width=0.45\textwidth]{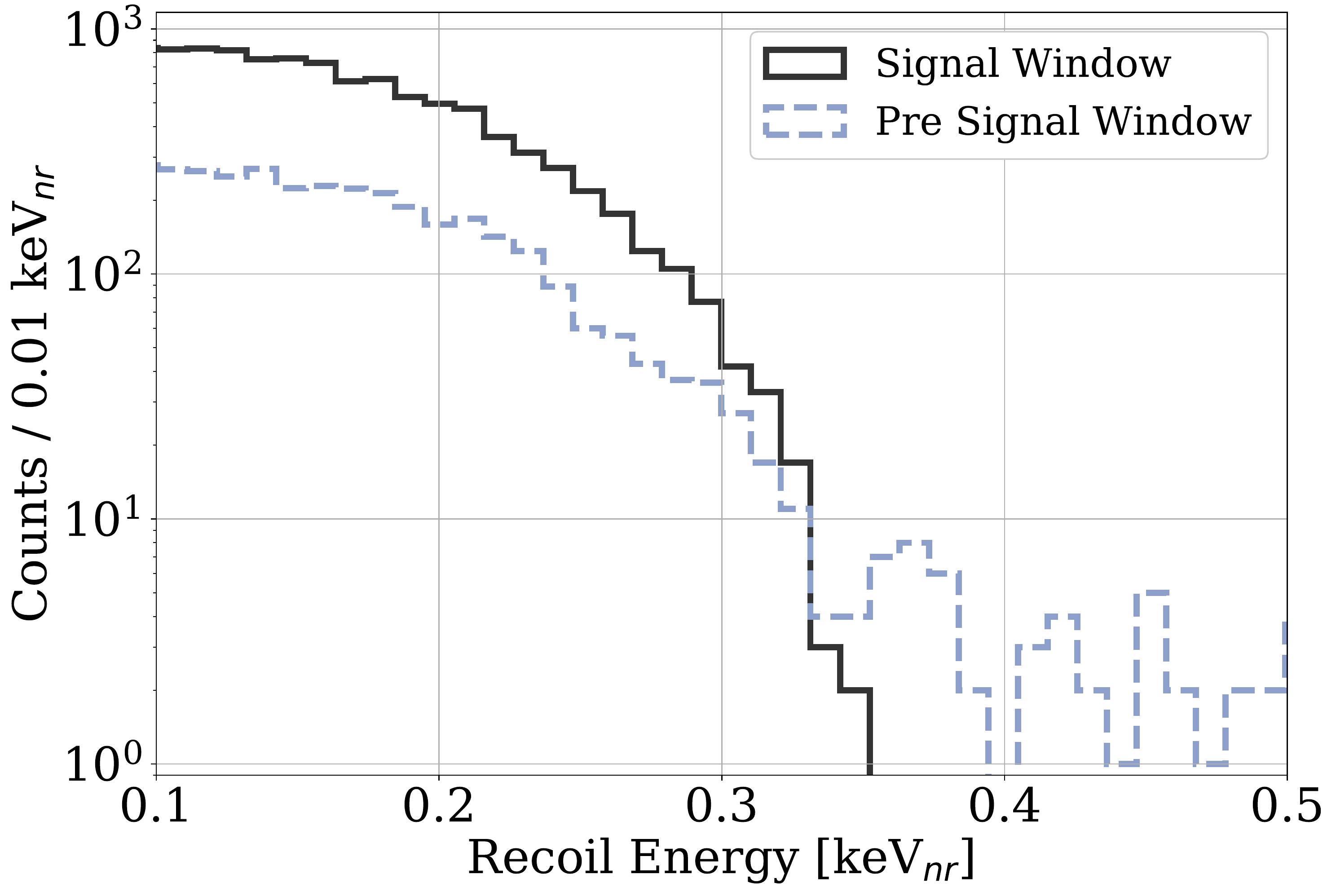}
    \caption{Recoil energy distributions of the signal events inside the signal window (solid) and before the signal window (dashed) for $w_n=30\,\mu$s and a 5\,cm water tank moderator. 
    Waiting until the last scatter occurs ensures that the capture of fast neutrons, which are associated with larger recoil energies, are not included in the analysis.}
    \label{fig:timedRecoil}
\end{figure}

Although the time distribution of neutron capture events is relatively unaffected by the pulse width, the time of last scatter and therefore the number of capture events that occur in the signal window is sensitive to $w_n$.
$\text{N}_\text{signal} (E_{\downarrow}, w_n)$ is defined as a measure of the fraction of usable signal events, such that
\begin{align} \label{eq:SR_NR2}
    &\text{N}_\text{signal} (E_{\downarrow}, w_n)= \nonumber \\
    &\frac{\text{Number of capture signals after the last scatter}}{\text{Total number of capture signals}},
\end{align}
where as in Eq.~\ref{eq:SR_NR}, only NR deposits with energy less than $E_\downarrow$ are kept.
Figure~\ref{fig:timedRecoil} shows the recoil energy distributions of signal events before and inside the signal window for a 5\,cm water tank and $w_n = 30\,\mu$s.   
Excluding early signal events with a time cut has the benefit of removing events with extra recoil energies attributed to the faster neutron collisions.
This time cut retains a majority (80\%) of the signal events.

The metrics $\text{T}_\text{NR}$ and $\text{N}_\text{signal}$ summarize the general features of the neutron interaction time structure.
These are evaluated for different thicknesses of the water tank and shown in Table~\ref{tab:geometry} for $w_n = 30\,\mu$s.
The rate of neutron capture in the TPC drops for both small and large tanks.
For small tanks, the rate drops due to insufficient neutron moderation.
For large tanks, it is due to fewer neutrons making their way into the TPC.
However, larger tanks offer a greater degree of scatter-capture separation.

\begin{table}
    \centering
    \begin{ruledtabular}
    \begin{tabular}{c c c c c}
        Thickness [cm] & Captures [\%] & Signals [\%] & $\text{T}_\text{NR}$ & $\text{N}_\text{signal}$ \\
        \midrule
        2  & 0.007 & 0.0010 & 0.02 & 0.61  \\
        5  & 0.032 & 0.0047 & 0.15 & 0.81  \\
        10 & 0.045 & 0.0068 & 0.36 & 0.89 \\
        15 & 0.033 & 0.0051 & 0.57 & 0.92 \\
        20 & 0.020 & 0.0030 & 0.71 & 0.94 \\
        25 & 0.010 & 0.0016 & 0.98 & 0.95 \\
    \end{tabular}
    \end{ruledtabular}
    \caption{Properties that influence the choice of water tank moderator thickness, including the neutron capture and signal percentage of neutrons entering the water tank, the signal separability metric $\text{T}_\text{NR} (1 \text{ keV}_\text{nr}, 30\, \mu s)$, and $\text{N}_\text{signal} (1 \text{ keV}_\text{nr}, 30\, \mu s)$.}
    \label{tab:geometry}
\end{table} 

\subsection{Neutron Pulse Width} \label{ss: pulseWidth}

At first glance, it may appear advantageous to have a large pulse width by considering the proportional increase in neutrons emitted per pulse.
However, signal events are selected using the signal window time cut, which has an efficiency that depends on the pulse width.
Since the signal window is defined to be between when the neutron scattering and neutron capture events end, the timing of the capture and scatter processes are analyzed as a function of pulse width.

The time structure of the neutron capture population does not have a significant dependence on $w_n$.
The thermalization process in the water tank sets a characteristic time scale of $\mathcal{O}$(100\,$\mu$s) for the neutron capture distribution (see Fig.~\ref{fig:timeStructure} for the 5\,cm tank). 
As long as this time scale is greater than $w_n$, the time structure of the captures is insensitive to changes in $w_n$.
By the same argument it is noted that the neutron scattering population is more responsive to changes in $w_n$ because scatters occur much earlier than the bulk of neutron captures.
As an example, the characteristic scattering time set by a 5\,cm water tank is around 10\,$\mu$s, and thus the resulting time structure is affected by values of $w_n$ larger than 10\,$\mu$s (in Fig.~\ref{fig:timeStructure} the last scatter time is prolonged to 30\,$\mu$s).
While the signal window shrinks as its beginning is postponed with increasing $w_n$, the reduction is negligible until $w_n$ approaches the time scale of neutron captures. 

\begin{figure}
    \centering
    \includegraphics[width=0.45\textwidth]{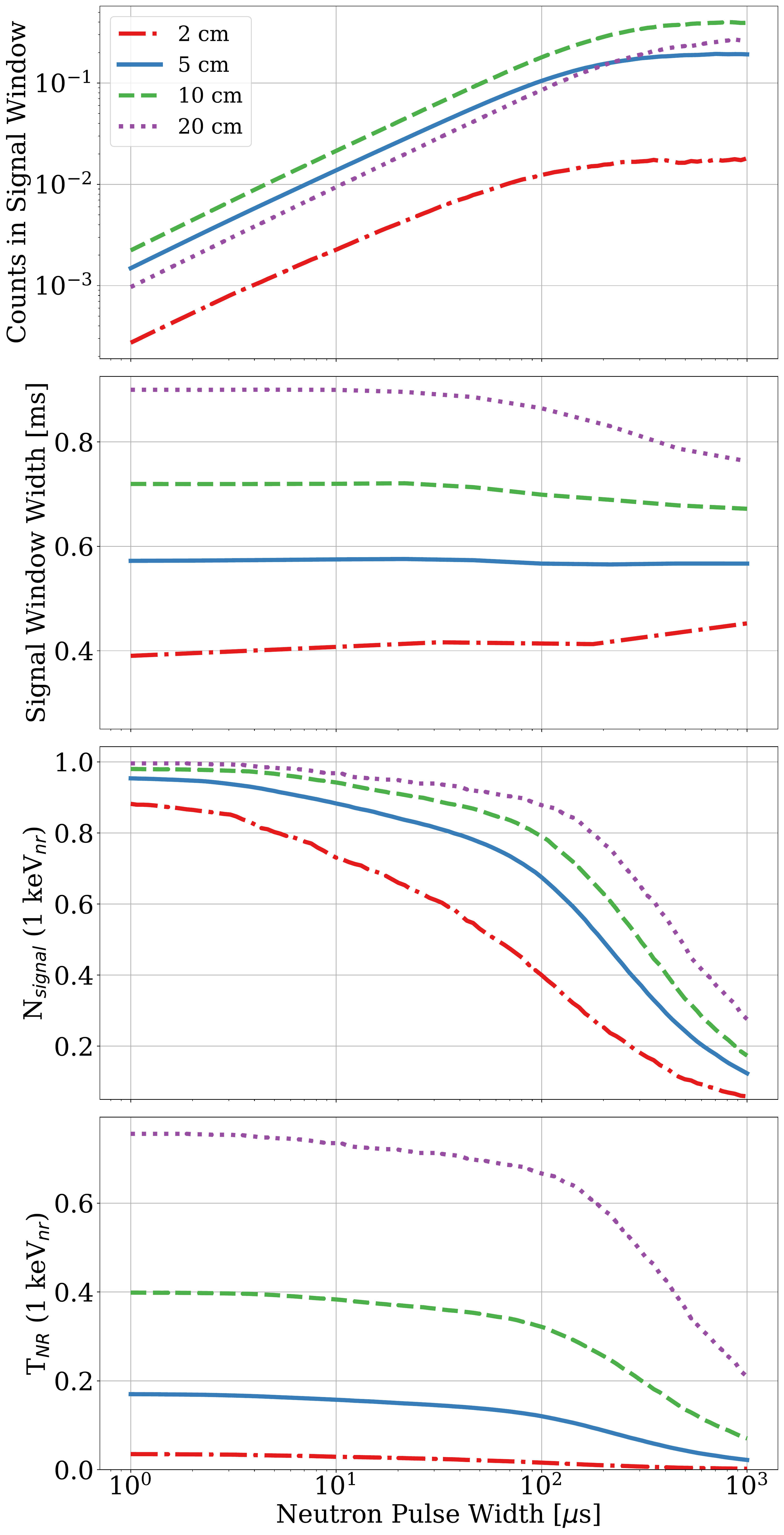}
    \caption{Simulated metrics as a function of neutron pulse width for 10$^9$ n/s and various water tank thicknesses. 
    \emph{Top:} Number of signal events falling inside a signal window.
    \emph{Center top:} Width of the signal window, which begins after the last scattering event and ends when 99\% of signal events have been produced after the last scatter. 
    \emph{Center bottom:} $\text{N}_\text{signal}$ for events that deposit less than 1 keV$_\text{nr}$ in the TPC.
    \emph{Bottom:} $\text{T}_\text{NR}$ for events that deposit less than 1 keV$_\text{nr}$ in the TPC.
    }
    \label{fig:pulseWidth}
\end{figure}

The results are summarized in the top two panels of Fig.~\ref{fig:pulseWidth}.
The first panel shows the number of signal events in the signal window as a function of neutron pulse width for various water tank thicknesses.
The proportional increase tapers off when $w_n$ approaches the neutron thermalization time set by the water tank, because the signal window begins at later and later times as the scattering events are prolonged.
The second panel confirms that the scattering and capture distributions, which set the beginning and end of the signal window, are insensitive to changes in the pulse width until it is on the order of the thermalization time in each tank.
These considerations suggest that using an arbitrarily large neutron pulse width would be beneficial, if not for the degradation of the separability metrics $\text{N}_\text{signal}$ and  $\text{T}_\text{NR}$, defined in Eqs.~\ref{eq:SR_NR} and~\ref{eq:SR_NR2}.
These quantities are shown as a function of $w_n$ in the bottom two panels of Fig.~\ref{fig:pulseWidth}. 
The fall of these metrics at longer pulse widths is caused by the extension of the last scatter time.

As a result the neutron pulse width is constrained to be no more than $\mathcal{O}$($100 \, \mu$s).
The generalization of this constraint can be obtained by comparing the characteristic timing of scattering and capture processes in a detector.
A stronger constraint on the pulse width arises when considering the mitigation of background and puile-up events produced directly by neutron captures that could pollute the signal window. 
This is further discussed in Section~\ref{Background}.

\section{Expected Backgrounds} \label{Background}

We now consider three types of non-NR events that could reduce usable signal counts:
i) the low energy ER background from the $\gamma$ cascades of activated and capture products, and from radiation in the environment,
ii) the single electron (SE) background, and
iii) the high energy ER events in the TPC. 
The first two produce small S1 and S2 signals that may overlap the faint signature of neutron capture events. 
In contrast, the third produces large S1 and S2 signals that may coincide with the signal events in time, temporarily blinding the detector.

\subsection{Low Energy ER Background} \label{ER background}

Neutrons in the vicinity of the detector are an indirect source of ER events in the TPC due to the de-excitation cascades of nuclei that undergo neutron interactions. 
$\gamma$-producing neutron interactions (capture or inelastic scatter) can happen both inside and outside the TPC.
Most of the ER events in the MiX detector originate from outside the TPC, where there are large amounts of LXe and water (see Fig.~\ref{fig:MiX}). 
These events are called external ER events, as opposed to internal ER events that are accompanied by a small NR signature.
Table~\ref{tab:eventClassify} shows neutron capture events partitioned according to where the capture and subsequent ER energy deposit occur.

\begin{table}[h]
    \centering
    \begin{ruledtabular}
    
    \begin{tabular}{c|c|c|c}
        \multicolumn{2}{c}{} & \multicolumn{2}{c}{\textbf{ER Deposit}} \\
        \cmidrule{3-4}
        \multicolumn{2}{c}{} & Inside TPC & Outside TPC \\
         \cmidrule{2-4} 
         \multirow{2}{5em}{\textbf{Neutron Capture}} & Inside TPC & Internal Bkgd. & Signal \\
         \cmidrule{2-4} 
         & Outside TPC & External Bkgd. & Undetected \\ 
    \end{tabular}
    \end{ruledtabular}
    \caption{Classification of events based on where the $\gamma$-producing neutron interaction and subsequent ER energy deposit took place.}
    \label{tab:eventClassify}
\end{table}

\begin{figure}
    \centering
    \includegraphics[width=0.45\textwidth]{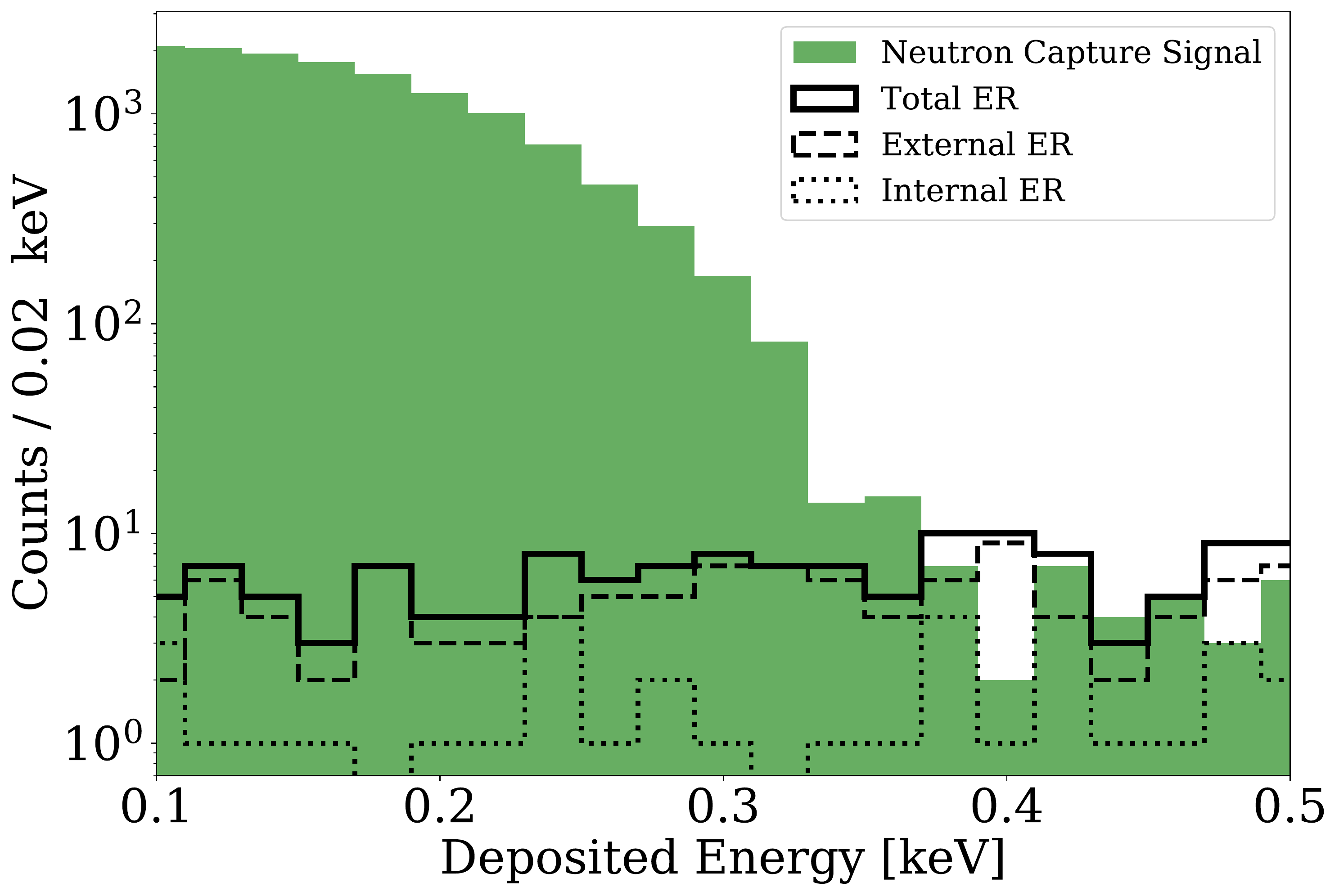}
    \caption{Deposited energy spectrum due to the internal (dotted) and external (dashed) ER background below $0.5$ keV without clustering applied, for a 5\,cm water moderator. 
    Also shown is the corresponding recoil spectrum due to the neutron capture signals.
    This simulation assumes a 1.2 day exposure with a 30\,$\mu$s pulse width and 60~Hz pulsing frequency, resulting in 20,000 neutron capture signal events.
    The number of ER counts below 0.5 keV is less than 0.1\% of the number of NR signal counts.}
    \label{fig:ERBackground}
\end{figure}

The low energy component of both the internal and external ER events are found to be small compared to the number of neutron capture signal events for the tank thicknesses considered here.
Figure~\ref{fig:ERBackground} shows the internal and external components of the ER background for a 5\,cm water moderator without clustering applied to the energy deposition sites. 
This represents an upper bound of the ER counts, amounting to less than 0.1\% of the number of neutron capture signals below 0.5 keV.
As expected for a large volume of LXe outside the TPC, most of the background is external, and a large majority (95\%) originate from neutron capture $\gamma$ cascades.

\subsection{Single Electron Background} \label{single e background}

Small electron backgrounds are one of the biggest obstacles to the low energy sensitivity of LXe TPCs.
Their high rate poses challenges to searches where the expected ionization signal is only a few electrons, as for ionization-only analyses, or in searches for the coherent scattering of solar neutrinos~\cite{angle2011search, aprile2016low, aprile2019light, akerib2019results, aprile2021search}. 
Single electron (SE) backgrounds are particularly challenging for low energy yield measurements because a significant fraction of NR events below 0.3 keV$_\text{nr}$ produce only one electron~\cite{lenardo2019low}.
According to the NEST model, of all neutron capture signals that produce an ionization signal, 80\% produce a single electron.

Although the origin of the SE background is not known with certainty, it has been observed that it is almost always preceded by large ionization signals.
Background SE events have been observed to persist much longer than the maximum drift time after the initial interaction~\cite{kopec2021correlated, akerib2020investigation}.
Further, this time behavior has been found to depend on runtime parameters like the purity of the LXe and the magnitude of the electric field~\cite{sorensen2018two}.
Despite the dedicated studies that have been performed using data from multiple detectors, an accurate simulation of this background is still out of reach~\cite{akerib2020investigation, aprile2014observation}. 

The rate of the SE background is expected to be higher in the signal window, due to the capture-induced $\gamma$ cascades adding on to $\gamma$ radiation from the environment. 
The background SE events are indistinguishable from electrons produced by capture-induced recoils of signal events.
They will have to be subtracted following a measurement of the SE rate in the signal window, in a manner similar to the background subtraction in Ref.~\cite{lenardo2019measurement}.
Due to difficulties in modeling the SE background, it can only be properly addressed after an explicit measurement.
After such a measurement there are solutions for mitigation at the hardware~\cite{sorensen2017electron, akerib2020investigation} and analysis~\cite{akeribImproving} levels.

The background SE rate must be taken into account when deciding the neutron pulsing frequency. 
Given that all neutron interactions following a 30\,$\mu$s pulse die off 1\,ms after the start of the pulse (see Fig.~\ref{fig:timeStructure}), a strict upper bound on the pulsing frequency can be set at 1\,kHz.
Above this frequency, neutron scatters will start to overlap the isolated capture population. 
In practice it is likely that high background SE rates in the signal window will disfavor the maximum pulsing frequency, and that some time is needed after a pulse for the background SE rate to decay away. 
A trade off will have to be made on the pulsing frequency to optimize the number of background-subtracted single electrons produced by signal events.
For this study, a pulsing frequency of 60\,Hz is assumed based on an investigation into the decay rate of the SE background performed by the LUX experiment, where the intensity of the SE rate was observed to drop ten-fold in 16\,ms~\cite{akerib2020investigation}.
The pulsing frequency will have to be tuned following a measurement of the SE rate and its decay constant in the MiX detector.

\subsection{Pileup From High Energy ER Events} \label{subsec: HE-ER}

\begin{figure}
    \centering
    \includegraphics[width=0.45\textwidth]{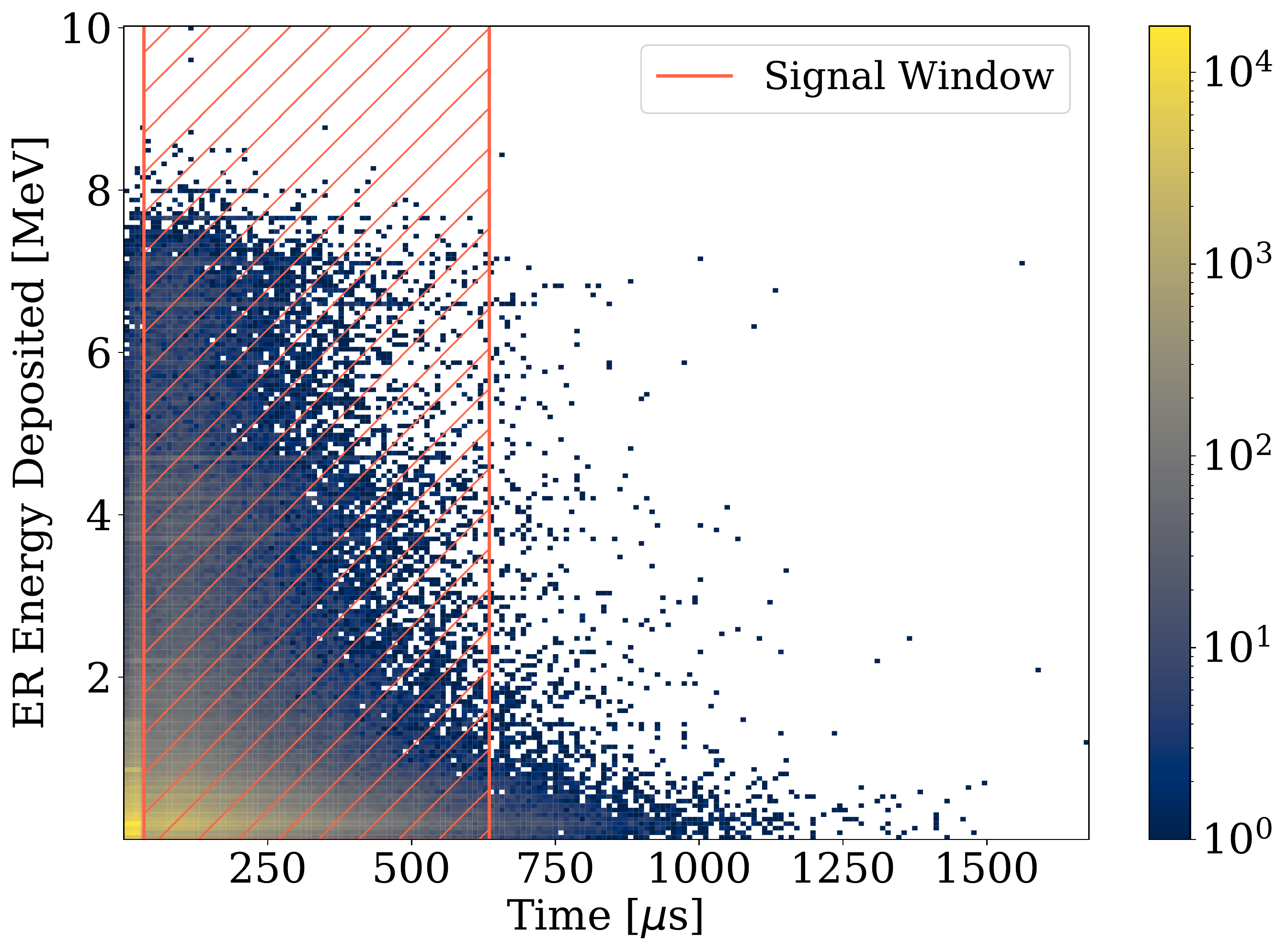}
    \caption{Distribution of high energy ER deposits in the TPC as a function of time elapsed since the beginning of a neutron pulse of width 30\,$\mu$s. Simulated for a 5\,cm water tank, ER events resulting from 3,000 pulses are shown, corresponding to about 12 signal events. Shown in orange is the signal window for this configuration, presented in the center top panel of Fig.~\ref{fig:pulseWidth}. 
    }
    \label{fig:highER}
\end{figure}

High energy ER events can coincide with the neutron capture signal and contribute to pileup, decreasing the number of clean acquisition windows that contain only the S1 and S2 pulses of the signal event.
The source of these ER events can be both internal and external, although in the MiX detector the internal component is negligible. 
A distribution of ER energy deposits in time, summed over many neutron pulses, is shown in Fig.~\ref{fig:highER} for a 5\,cm water tank.
The two main contributors to this background are the hydrogen in the water and the LXe outside the TPC.
In the following, the ER pileup is minimized and constraints for the size of the water tank and length of the neutron pulse are obtained.

\begin{figure}
    \centering
    \includegraphics[width=0.47\textwidth]{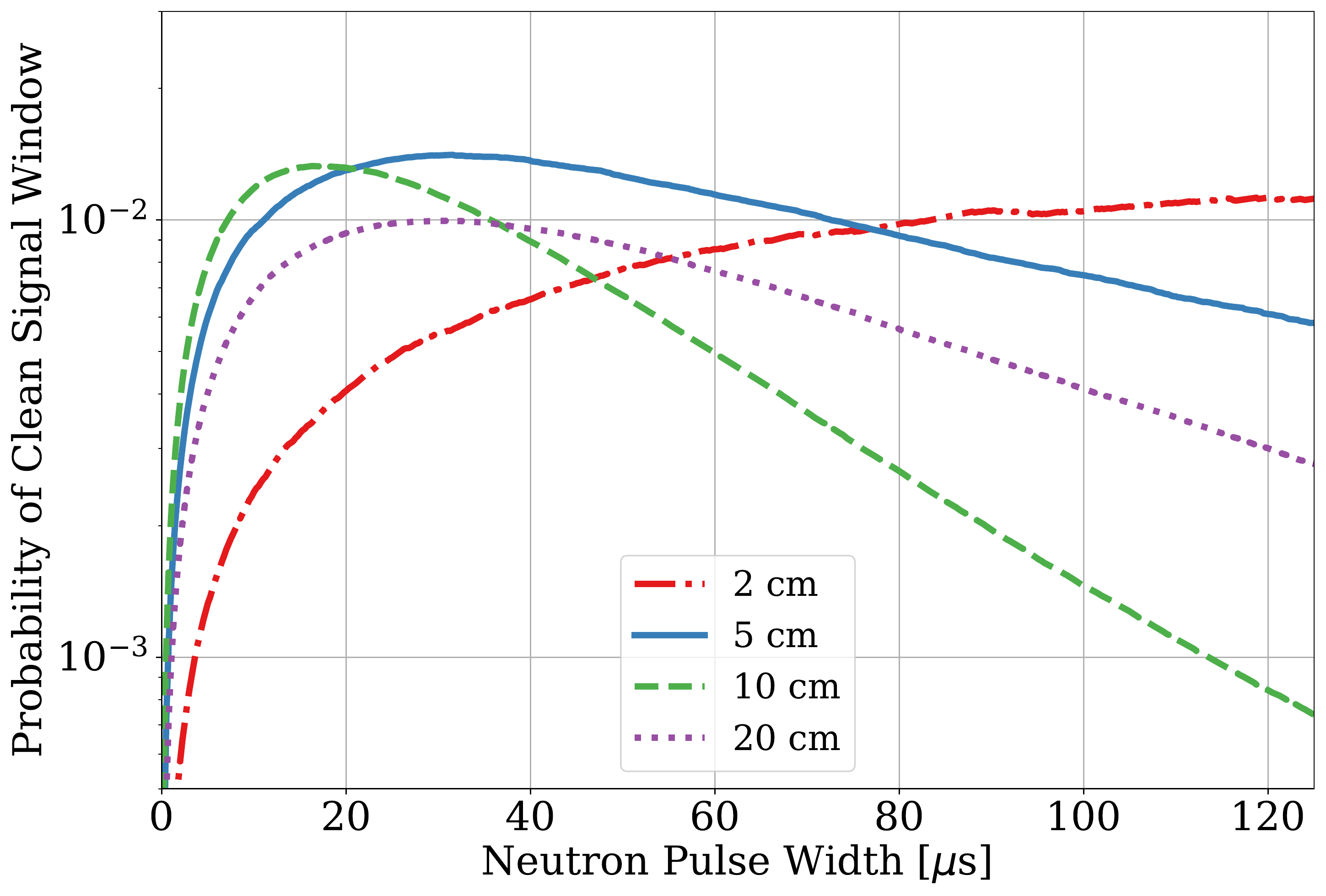}
    \caption{Probability $P$ of obtaining a clean signal window where no signal event is accompanied by an ER deposit, as a function of neutron pulse width $w_n$. 
    Curves are shown for a representative set of water tank thicknesses.
    }
    \label{fig:cleanSignal}
\end{figure}

Within a given signal window, both the number of high energy ER events originating from neutron capture and the number of neutron capture signals are modeled according to Poisson distributions. 
This is a valid approach as long as neutron-induced interactions from previous pulses do not leak into the current signal window. 
As mentioned in Section~\ref{single e background}, this leakage would only occur for pulsing frequencies larger than 1\,kHz.
Each pulse can be treated as independent.
The quantity of interest is the probability $P$ that a given signal window has no large ER deposits, while containing a signal event, such that
\begin{equation}
    P = \text{Pois}(0, \text{ER}_\text{external}) \times \text{Pois}(1, \text{NR}_\text{signal}),
\end{equation}
where $\text{ER}_\text{external}$ is the average number of external ER events in a signal window, and $\text{NR}_\text{signal}$ is the average number of signal events in a signal window.
$\text{NR}_\text{signal}$ automatically excludes internal ER contributions since signal events are defined as captures inside the TPC that are not accompanied by ER deposits.
Figure~\ref{fig:cleanSignal} shows the probability of a given signal window having one signal event and no ER deposits as a function of $w_n$ for several thicknesses of the water tank moderator. 
While the number of signal events proportionally increases with $w_n$, there is a corresponding rise in the rate of external ER background due to the capture products of material outside the TPC.
This sets a strong constraint on the optimal value for the neutron pulse width, specific to each water tank thickness.
A water tank with thickness 5\,cm and $w_n = 30\,\mu$s are identified as optimal.
Note that the ER pileup can be further mitigated if the capture signal is tagged by its $\gamma$ cascade using the LXe skin.
This allows to precisely determine the time when the capture occurred.

\section{Implications for Dark Matter Searches} \label{discussionSection}

\begin{figure}
    \centering
    \includegraphics[width=0.48\textwidth]{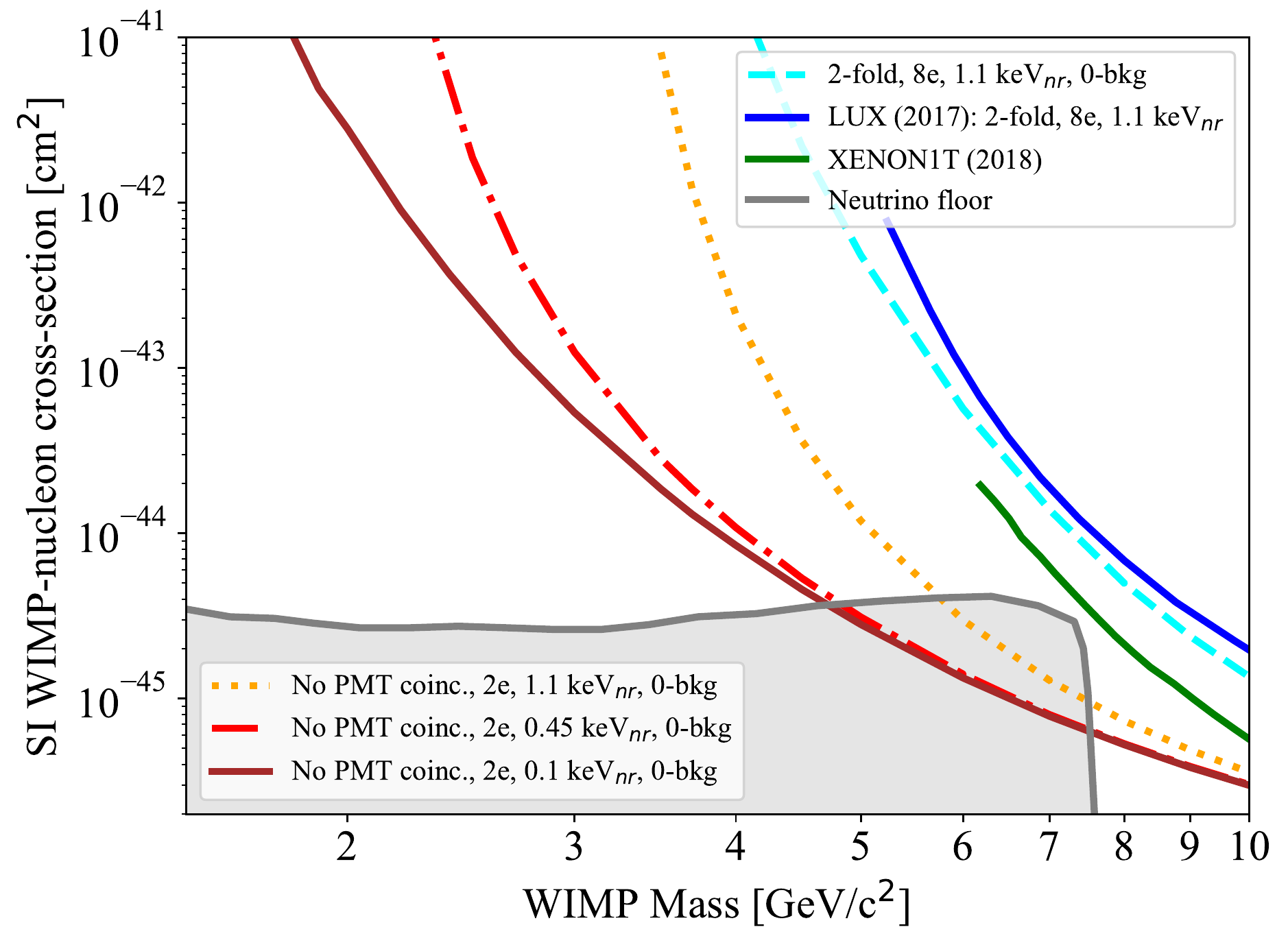}
    \caption{The projected $90\%$ sensitivities for a generic background-free LXe experiment with a full LUX-like exposure are shown for different energy thresholds in solid maroon, red, and orange curves.
    The limits are generated using the NEST 2.0.1 default yield models~\cite{szydagis_m_2019_3357973}. 
    The searches use both ionization and scintillation channels with no PMT coincidence requirement and a two extracted-electron threshold. 
    A $0\%$ signal acceptance is enforced for recoil energies below the indicated values.
    The solid and dashed blue curves verify that the LUX result is fairly reproduced~\cite{run4fullexpo}.
    Also shown are limits from LZ (dashed black)~\cite{lz2022}, XENON1T (dashed green)~\cite{Aprile:2018dbl}, and DarkSide-50~\cite{abdelhameed2019first} (dashed purple)}.

    \label{fig:limit}
\end{figure}

Measurements of the NR quanta yields in LXe below $0.3$ keV$_\text{nr}$ would provide an absolute detector-independent calibration for LXe experiments.
A lower energy threshold allows xenon interactions with slower WIMPs to be detected, increasing the number of observable WIMP events.
The increase in counts becomes more significant for lighter WIMPs, where the cutoff velocity for particles to produce detectable NR events is in the tail of the Maxwell-Boltzmann velocity distribution.
Figure~\ref{fig:limit} shows the gain in sensitivity for light WIMPs in an idealized LXe detector, assuming the yield models in NEST~v2.0.1 (which was modified to remove the sharp cutoff in the yields at 0.2\,keV$_\text{nr}$) are experimentally realized down to $0.1$ keV$_\text{nr}$~\cite{szydagis_m_2019_3357973}.
These sensitivity curves assume a two extracted-electron threshold, a $0\%$ WIMP acceptance for recoil energies below various energy thresholds, and no PMT coincidence required in NEST. 
In addition to greater sensitivity to light WIMPs, lower energy thresholds offer the following benefits.
First, if light (below $10$\,GeV) WIMPs are discovered, the interaction cross section can be reconstructed with higher precision~\cite{peter2014wimp}.
The interaction cross section for light WIMPs suffers a degeneracy because only the high-$v$ tail of the velocity distribution is probed.
Furthermore, performing an ultra-low energy NR calibration will allow the routine projections down to 0.1\,keV$_\text{nr}$ found in the direct detection literature to be either corroborated or refuted~\cite{akerib2020projected, gelmini2018casting}.

\section{Conclusion} \label{conclusion}

This study focuses on an experimental concept to measure the LXe response to low energy NR interactions using the recoils of neutron capture products.
It is emphasized that using a small detector with high light and charge collection efficiencies enhances the chances of a successful measurement.

To establish feasibility, simulations were performed for the MiX detector, a small dual phase TPC designed to maximize light collection, and a pulsed D-D generator as a neutron source.
The small size of the active LXe volume allows about 15\% of the $\gamma$ cascades resulting from neutron capture to escape the TPC, leaving behind a pure NR signature.
A pulsed neutron source induces a time structure for the neutron interactions that allows a large fraction of the neutron capture events (80\%) to be isolated.
These signals can be positively identified using an instrumented LXe volume outside the TPC that can record the $\gamma$ cascades. 
The isolated fraction is found to depend on the thickness of the water tank moderator that surrounds the detector, and a trade off has to be made between the higher statistics allowed by smaller tanks, and the higher separability of signals allowed by larger tanks.

The parameters of the neutron generator are found to affect the numbers of signal, background, and pile-up events.
A trade off also has to be made between a large neutron pulse width, which increases the number of neutron capture events, and a small pulse width, which decreases the rate of ER pile-up originating from neutron capture outside the TPC.
A similar but independent compromise is struck for the pulsing frequency, which is constrained from above to mitigate single electron backgrounds, but needs to be sufficiently high for a proper background subtraction.

The neutron capture population identified in this study constitutes an ideal set of events for probing the scintillation and ionization yields down to 0.13\,keV$_\text{nr}$, with the recoil events at 0.3\,keV$_\text{nr}$ serving as a cross-reference to the current lowest measured ionization yield~\cite{dqthesis}. 
Whether or not the fundamental limits of NR quanta production are accessible, the results of such a measurement will extend the present knowledge of low energy physics in LXe, and increase the power of direct detection experiments that use it.

\begin{acknowledgments}
We would like to thank M. Solmaz, M. Szydagis, P. Sorensen, R. Linehan, J. Liao, and X. Xiang for insightful conversations. This research was supported by the Department of Energy under Grant SC0019193, and with resources provided by the Open Science Grid~\cite{pordes2007open, sfiligoi2009pilot}, which is supported by the National Science Foundation award \#2030508.
\end{acknowledgments}

\appendix

\section{Neutron Capture Model Uncertainty} \label{app:A}

The GEANT4 photon evaporation model simulates $\gamma$ cascades by  sampling the evaluated experimental nuclear structure data (ENSDF) maintained by the national nuclear data center at Brookhaven National Laboratory~\cite{geant4manual, geant4photonevap}. 
The model then uses the $\gamma$ energy spectra and multiplicity distributions as inputs to generate the neutron capture recoil events.
To the best of our knowledge, the $\gamma$ energy spectra for each multiplicity and the multiplicity distributions have not been measured for any isotope of xenon except $^{136}$Xe~\cite{albert2016measurement}.
A custom model to generate recoil events was written using the $\gamma$ spectra and multiplicities as input parameters to study the effect on the NR spectrum.
The sources of uncertainty arising from those parameters and the sampling method in the custom model were combined to obtain an uncertainty of the GEANT4 NR spectrum.
A direct comparison of GEANT4 and data from $^{136}$Xe is also made.

\subsection{NR uncertainty calculation}

The evaluated gamma ray activation file (EGAF) is a database of neutron capture $\gamma$ ray energies and cross sections prepared by the International Atomic Energy Agency~\cite{EGAFdata}. 
This database was formed by merging elemental $\gamma$ spectrum measurements taken in 2002 at the Budapest Research Reactor using a high-purity germanium detector with nuclear structure data~\cite{EGAFarticle}. 
By comparing the GEANT4 spectra with the EGAF database, the uncertainty in the NR spectrum was estimated.

Since the EGAF database does not include multiplicity information, it is not possible to adjust the photon energies in GEANT4 and repeat the simulation. 
Instead, a model is constructed to produce recoil events from a single, multiplicity-independent photon energy spectrum for each isotope.
The recoil events generated by the model can be combined according to isotope abundances and an estimate for the multiplicity distributions to produce a final NR spectrum.
The algorithm is as follows:
Given a normalized photon energy spectrum, neutron separation energy S$_\text{n}$~\cite{Gammabook}, and desired multiplicity $\kappa$, take $\kappa$-1 random samples from the spectrum and calculate the sum of their energies $E_{\kappa - 1} = \sum_{i=1}^{k-1} E_i$. 
Let E$_\text{min}$ and E$_\text{max}$ be the respective lowest and highest photon energies in the sampled spectrum.
Then if S$_\text{n}$~-~E$_\text{max}$~$<$~$E_{\kappa - 1}$~$<$~S$_\text{n}$~-~E$_\text{min}$, set the final $\gamma$ energy to S$_\text{n}$ - $E_{\kappa - 1}$ to conserve energy. 
If $E_{\kappa - 1}$ is not in the acceptable range, reject the event and re-sample a new set of $\kappa -1$ energies. 
Once a complete set of $\kappa$ photons has been generated, choose random directions in 4$\pi$ for each $\gamma$ and calculate the nuclear recoil using momentum conservation.

\begin{figure}[h]
    \centering
    \includegraphics[width=0.48\textwidth]{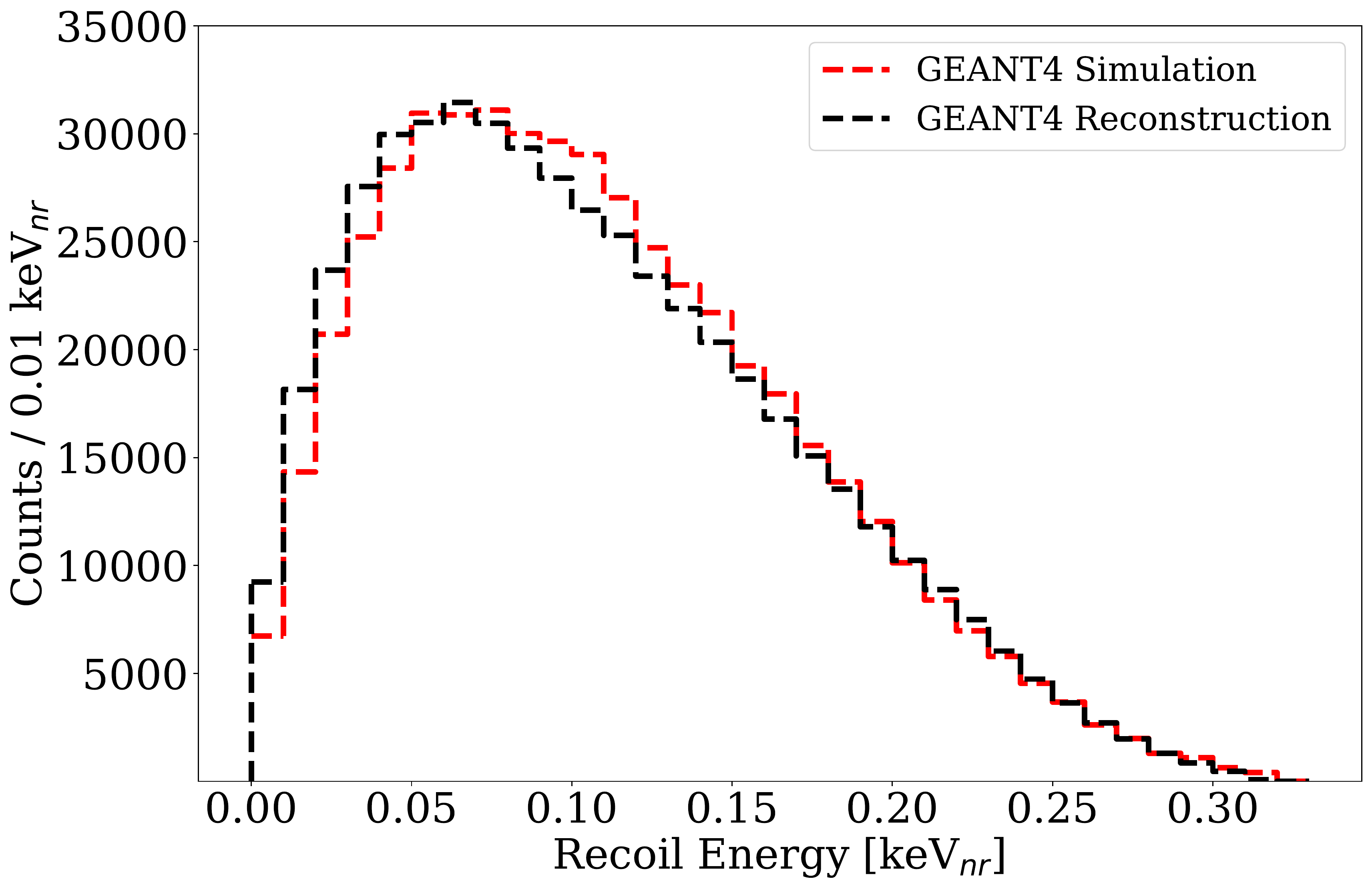}
    \caption{Nuclear recoil spectra for 500,000 neutron capture events produced by the GEANT4 simulation and reconstruction for $^{129}$Xe, $^{130}$Xe, and $^{131}$Xe. 
    The reconstructed recoil energies skew slightly lower than the GEANT4 simulation, but the spectra match well above 0.13\,keV$_\text{nr}$.
    }
    \label{fig:modelcomp}
\end{figure}

To compare the GEANT4 results with the EGAF database, the uncertainty associated with the sampling method must be accounted for. 
The simulation provides multiplicity-independent $\gamma$ energy spectra for each isotope.
The average multiplicity $\kappa_\text{avg}$ is calculated by dividing S$_\text{n}$ by the average $\gamma$ energy. 
Although it is possible to directly extract multiplicities from GEANT4, a Gaussian distribution about $\kappa_\text{avg}$ is assumed so that the same process may be applied to the EGAF data, which does not contain multiplicity information. 
The effective cross section is calculated for each isotope as the product of the natural abundance and the thermal neutron capture cross section. 
The effective cross sections are used to weigh the fraction of samples taken from each isotope’s energy spectrum. 
The effective cross sections indicate that $^{129}$Xe, $^{130}$Xe, and $^{131}$Xe make up more than 99\% of neutron capture events.
Thus the analysis is restricted to these three isotopes. 
With a total of 500,000 events, the sampling method is used to generate recoils, and the result is compared with recoils simulated by GEANT4. 
As shown in Fig.~\ref{fig:modelcomp}, the recoil spectra match well above 0.13\,keV$_\text{nr}$, validating the sampling method and the multiplicity assumption.

\begin{figure}
    \centering
    \includegraphics[width=0.48\textwidth]{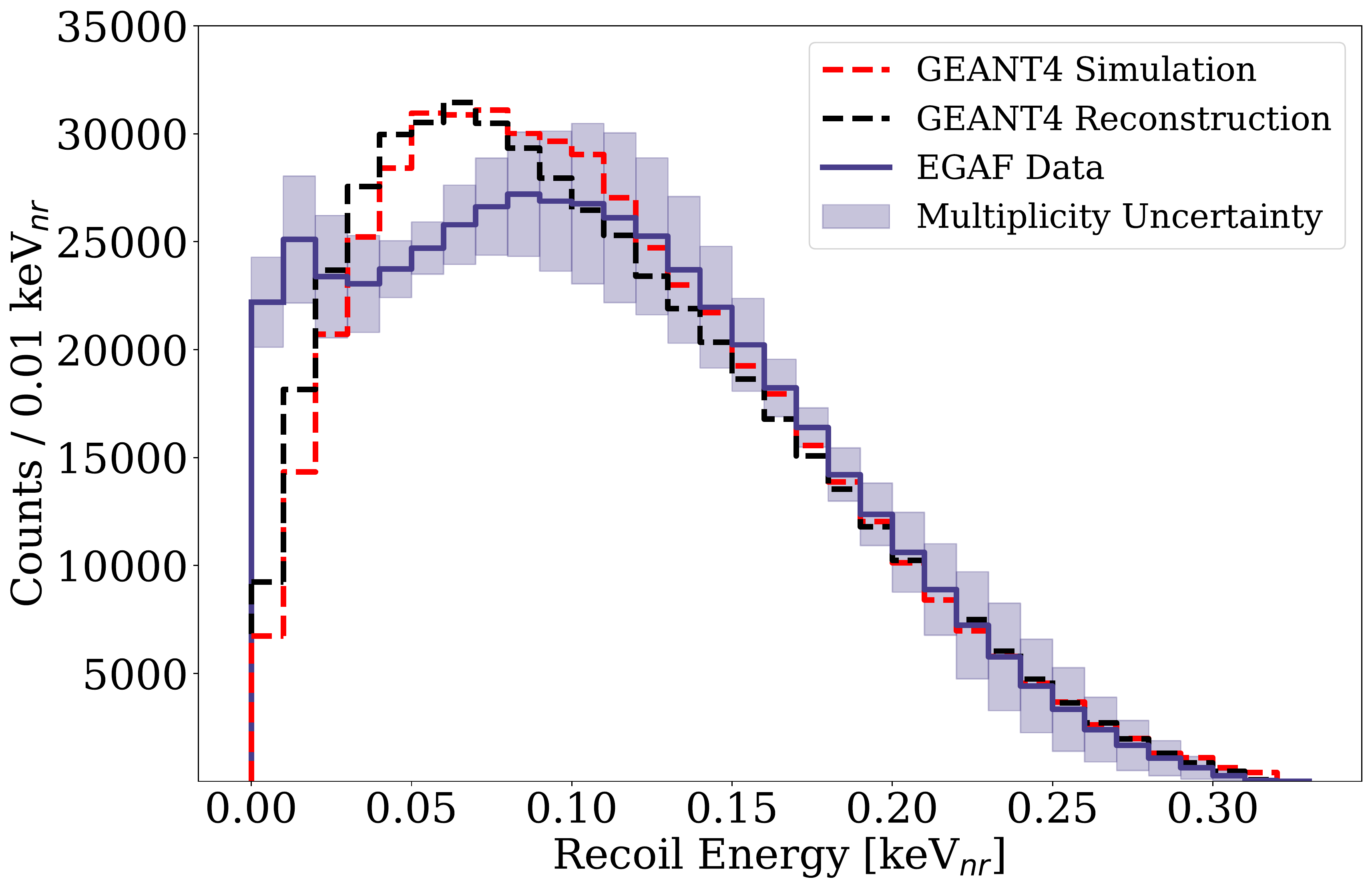}
    \caption{Nuclear recoil spectra for 500,000 neutron capture events produced by GEANT4 simulation, GEANT4 reconstruction, and reconstruction using the EGAF database. The uncertainty in the EGAF NR spectrum (purple band) is calculated by varying the width of the Gaussian $\gamma$ multiplicity distribution for each isotope. 
    The EGAF reconstruction matches the GEANT4 simulation closely for energies greater than 0.13\,keV$_\text{nr}$.
    }
    \label{fig:egafcomp}
\end{figure}

Next, the same sampling procedure is carried out using $\gamma$ energy spectra from the EGAF database to compare with the NR spectrum obtained from GEANT4. 
A uniform standard deviation of 1.5 was chosen for the Gaussian multiplicity distributions for the $\gamma$ spectra obtained from GEANT4, because that value produced the recoil spectrum that most closely matched the GEANT4 simulation. 
For EGAF, however, there exist no multiplicity distributions. 
For each isotope, the EGAF data are sampled using Gaussian distributed multiplicities with standard deviations between 1 and 5. 
The absolute minimum and maximum counts of NR events are calculated in each energy bin.
Those counts become the lower and upper bounds for the EGAF NR spectrum, shown in Fig.~\ref{fig:egafcomp}. 
The EGAF spectrum matches the GEANT4 simulation well for energies above 0.13\,keV$_\text{nr}$. 
The discrepancy at low energies is attributed to disagreements between the GEANT4 and EGAF $\gamma$ energy spectra for $^{130}$Xe.

The three sources of uncertainty, which include i) the uncertainty associated with simplifying assumptions made by the sampling method, ii) the discrepancy between the $\gamma$ spectra obtained from the EGAF database and the GEANT4 simulation, and iii) the variability in the EGAF NR spectrum calculated from varying the widths of the Gaussian multiplicity distributions, are added in quadrature to produce the final uncertainty band on the NR spectrum, shown in Fig.~\ref{fig:nCapError}. 
Ultimately, the uncertainty in the NR spectrum will be propagated to the yield models following the neutron capture calibration in the MiX detector.\\

\subsection{Comparison with \texorpdfstring{$^{136}$Xe data}{136Xe data}}

A similar analysis was performed for $^{136}$Xe using measurements of neutron capture $\gamma$ cascades taken at the Detector for Advanced Neutron Capture Experiments (DANCE) at the Los Alamos Neutron Science Center in 2016~\cite{albert2016measurement}. 
Unlike EGAF, these data include $\gamma$ energy spectra for each multiplicity as well as the overall multiplicity distribution. 
Since the maximum recoil energy of $^{136}$Xe is 60\,eV$_\text{nr}$, which is well below the target energy threshold, it will not contribute to the neutron capture calibration.
However, it is the only isotope that allows a comparison between the NR spectrum simulated in GEANT4 and a NR spectrum calculated from measured $\gamma$ spectra with multiplicity information.

$^{136}$Xe has a relatively small capture cross section and natural abundance (see Table~\ref{tab:nGamma}), and contributes only 0.1\% of the neutron capture events in natural xenon. 
Therefore, GEANT4 simulations were run using isotopically pure $^{136}$Xe rather than natural xenon to extract both the multiplicity distribution and the $\gamma$ energy spectra. 
NR spectra were produced for each multiplicity, then combined according to the weights specified by the multiplicity distribution. 
The same analysis was performed using DANCE data, and the count-weighted relative difference was calculated in the resulting NR spectra. 
Taking into account the multiplicity weights from the GEANT4 simulation and the DANCE data eliminates the systematic uncertainties associated with sampling the same $\gamma$ spectrum irrespective of multiplicity, and the assumption of Gaussian multiplicity distributions.
This allows to more accurately quantify the impact of variability of the $\gamma$ energies on the resulting NR spectra.

The discrepancy between the $^{136}$Xe NR spectra produced using $\gamma$ energies from GEANT4 and DANCE is represented by a weighted average difference of 40\%, taken over the full 0 - 60 eV$_\text{nr}$ energy range.
Above a 30\,eV$_\text{nr}$ threshold, the weighted average difference drops to 27\%. 
These differences indicate a disparity between the GEANT4 photon evaporation model and experimental data, and that it is particularly pronounced at low NR energies. 
However, note that $^{136}$Xe is not representative of other xenon isotopes because the GEANT4 $\gamma$ spectra for $^{136}$Xe are sparse (15 lines total) compared to those of more abundant isotopes (more $500$ lines each for $^{129}$Xe and $^{131}$Xe).
The small number of $\gamma$ energies makes the resulting $^{136}$Xe NR spectrum sensitive to discrepancies between the $\gamma$ spectra from GEANT4 and DANCE.
The larger number of lines in the other isotopes are expected to lead to smaller differences in the NR spectra due to discrepancies in $\gamma$ energy distributions.

\clearpage


%

\end{document}